\documentclass[review,showpacs,preprintnumbers,amsmath,amssymb]{elsarticle}

\usepackage{lineno}
\usepackage[colorlinks,hyperindex]{hyperref}
\usepackage{graphicx}
\usepackage{float}
\usepackage{amsmath}
\usepackage{amsfonts}
\usepackage{perpage}
\usepackage{multirow}
\usepackage{array}
\usepackage{dcolumn}
\usepackage{bm}
\usepackage{tikz}

\newcommand\copyrighttext{%
  \footnotesize \textcopyright \ 2019. This manuscript version is made available under the CC-BY-NC-ND 4.0 license http://creativecommons.org/licenses/by-nc-nd/4.0/}
\newcommand\copyrightnotice{%
\begin{tikzpicture}[remember picture,overlay]
\node[anchor=south,yshift=10pt] at (current page.south) {\fbox{\parbox{\dimexpr\textwidth-\fboxsep-\fboxrule\relax}{\copyrighttext}}};
\end{tikzpicture}%
}

\newcommand{\nd}{\noindent}
\newcommand{\vet}[1]{\mathbf #1 }
\newcommand{\um}[1]{\mathrm{#1}}

\newcolumntype{L}[1]{>{\raggedright\let\newline\\\arraybackslash\hspace{0pt}}m{#1}}
\newcolumntype{C}[1]{>{\centering\let\newline\\\arraybackslash\hspace{0pt}}m{#1}}

\modulolinenumbers[5]

\journal{Journal of \LaTeX\ Templates}

\begin{document}

\begin{frontmatter}

\title{Accessing temperature waves: a dispersion relation perspective}

\author{Marco Gandolfi}
\address{Laboratory of Soft Matter and Biophysics, Department of Physics and Astronomy, KU Leuven, Celestijnenlaan 200D, B-3001 Leuven, Belgium}
\address{Interdisciplinary Laboratories for Advanced Materials Physics (I-LAMP) and Dipartimento di Matematica e Fisica, Universit{\`a} Cattolica del Sacro Cuore, Via Musei 41, 25121 Brescia, Italy}
\ead{marco.gandolfi@kuleuven.be}

\author{Giulio Benetti}
\address{Laboratory of Solid State Physics and Magnetism, Department of Physics and Astronomy, KU Leuven, Celestijnenlaan 200D, B-3001 Leuven, Belgium}
\address{Interdisciplinary Laboratories for Advanced Materials Physics (I-LAMP) and Dipartimento di Matematica e Fisica, Universit{\`a} Cattolica del Sacro Cuore, Via Musei 41, 25121 Brescia, Italy}

\author{Christ Glorieux}
\address{Laboratory of Soft Matter and Biophysics, Department of Physics and Astronomy, KU Leuven, Celestijnenlaan 200D, B-3001 Leuven, Belgium}

\author{Claudio Giannetti}
\address{Interdisciplinary Laboratories for Advanced Materials Physics (I-LAMP) and Dipartimento di Matematica e Fisica, Universit{\`a} Cattolica del Sacro Cuore, Via Musei 41, 25121 Brescia, Italy}

\author{Francesco Banfi}
\address{Universit{\'e} de Lyon, Institut Lumi{\`e}re Mati{\`e}re (iLM), Universit{\'e} Lyon 1 and CNRS, 10 rue Ada Byron, 69622 Villeurbanne cedex, France\corref{mycorrespondingauthor}}

\ead{francesco.banfi@univ-lyon1.fr}

\begin{abstract}
\copyrightnotice
In order to account for non-Fourier heat transport, occurring on short time and length scales, the often-praised Dual-Phase-Lag (DPL) model was conceived, introducing a causality relation between the onset of heat flux and the temperature gradient. The most prominent aspect of the first-order DPL model is the prediction of wave-like temperature propagation, the detection of which still remains elusive. Among the challenges to make further progress is the capability to disentangle the intertwining of the parameters affecting wave-like behaviour. This work contributes to the quest, providing a straightforward, easy-to-adopt, analytical mean to inspect the optimal conditions to observe temperature wave oscillations. The complex-valued dispersion relation for the temperature scalar field is investigated for the case of a localised temperature pulse in space, and for the case of a forced temperature oscillation in time. A modal quality factor is introduced showing that, for the case of the temperature gradient preceding the heat flux, the material acts as a bandpass filter for the temperature wave. The bandpass filter characteristics are accessed in terms of the relevant delay times entering the DPL model. The optimal region in parameters space is discussed in a variety of systems, covering nine and twelve decades in space and time-scale respectively.
The here presented approach is of interest for the design of nanoscale thermal devices operating on ultra-fast and ultra-short time scales, a scenario here addressed for the case of quantum materials and graphite.
\end{abstract}


\end{frontmatter}

\clearpage


\section{\label{Introduction}Introduction}
\nd
The validity of Fourier's law \cite{fourier1822theorie}, the milestone constitutive relation describing diffusive heat transport, is affected by a major pitfall - notably the lack of causality between the heat flux and the temperature gradient - 
which manifest on short time and length scales \cite{huberman2019} and/or at low temperatures \cite{ackerman1968}. Non-Fourier schemes are thus required when dealing with heat transport in micro- and nano-devices and with devices operating at cryogenic temperature and/or on ultrafast time-scales \cite{Cahill2003,Cahill2014,volz2016,Cimelli2016,tzou2014macro,Vermeersch2018}.\\
\indent Among the non-Fourier formulations of heat transport, the much celebrated macroscopic Cattaneo-Vernotte model (CV) \cite{cattaneo1948sulla,cattaneo1958form,vernotte1958paradoxes,vernotte1961some} assumes the heat flux sets in after a delay time $\tau_q>0$, following the onset of a temperature gradient. Coupling the CV model constitutive relation with energy conservation leads to the \textit{telegraph equation} for the scalar temperature field, which is of the hyperbolic type. In this mode, temperature propagates as a damped wave with finite velocity. The thermal wave concept proved useful in a variety of different contexts where thermal inertia plays a role \cite{Joseph1989_RMP}.\\
\indent As a further gereralization, the DPL model \cite{tzou1995unified, tzou1995generalized, tzou1995experimental} introduced the possibility for precedence switching between the heat flux and the temperature gradient. The constitutive equation reads:
\begin{equation}
\vet{q}\left(\vet{r},t+\tau_q\right)=-\kappa_T\ \nabla T \left(\vet{r},t+\tau_T\right).
\label{DPLM_law}
\end{equation}
where $\vet{q}$ is the heat flux, $T$ the temperature, $\kappa_T$ the Fourier's thermal conductivity, and $\tau_T$ and $\tau_q$ are positive delay times. For $\tau_T<\tau_q$ the temperature gradient is the cause and the heat flux the consequence, whereas for $\tau_q<\tau_T$ the heat flux is the cause and the temperature gradient is the consequence, thus switching precedence in the causality relation\footnote{Actually, the DPL model is exactly the same as the one-lag model: $\vet{q}\left(\vet{r},t+\tau\right)=-\kappa_T\ \nabla T \left(\vet{r},t\right)$, where $\tau= \tau_q-\tau_T$ and, obviously, their exact solutions are the same. Expanding the models in Taylor series in time to first-order though, the DPL model and the single-phase lag model yield two different constitutive equations, the latter being exactly that of the CV model. This is due to the difference in which the expansion is taken, i.e. one linear expansion around t of extent $\tau_{q}$-$\tau_{T}$ for the single-phase lag model vs two linear expansions around t of extent $\tau_{q}$ and $\tau_{T}$ respectively. We refer the reader to Reference \cite{Ordonez-Miranda2010_MRC} for a thorough explanation of this important point.}.
Upon first order expansion in time and coupling with energy conservation, the DPL model leads to a \textit{Jeffrey's type equation} for the temperature scalar field, which is parabolic in nature.
Although the third-order mixed derivative term, which will be explicitly addressed further on, formally destroys the wave nature of the equation (the wave equation is hyperbolic, whereas the Jeffrey's type temperature equation is parabolic), the solution may still bear, under a practical stand-point, "wave-like" characteristics \cite{tang1999wavy}. Temperature propagation may thus preserve coherence properties, in contrast to the classical idea of an incoherent temperature field. This is the long sought behavior for applications dealing with micro- and nano-scale applications and/or ultra-fast time scales \cite{huberman2019}.\\ 
\indent The macroscopic DPL model, in its first-order formulation, has the merit of encapsulating a variety of microscopic models of non-Fourier heat transport arising from different physical contexts. 
For instance, the phonon scattering model \cite{guyer1966a_PR}, based on the solution of the linearized Boltzmann equation for the phonon field, was developed to investigate heat waves in dielectrics at cryogenic temperatures and, most recently, was applied \cite{torres2018_PRApplied} to rationalise seminal experimental findings on ultrafast thermal transport at the nano \cite{hoogeboom2015,frazer2019} and micro-scale \cite{Johnson2013}. The two-temperature model, derived solving the Boltzmann equation for the phonon-electron interaction \cite{Qiu1993}, effectivly described the ultrafast thermal dynamics in thin-films subject to impulsive laser heating \cite{Brorson1987} and proved successful in interpreting a multitude of phenomena ever since. Heat conduction in two-phase-systems \cite{Wang2008}, granular materials \cite{Miranda2010_granular}, nanofluids \cite{Wang2009,Khayat2015} and in the frame of bio-heat transfer \cite{Li2018} may all be cast in the frame of the first-order DPL formulation. 
As a matter of fact, the specific microscopic physics may be lumped in the delay times $\tau_{q}$ and $\tau_{T}$. This fact, on one side supports the validity of the macroscopic model itself, on the other it furnishes a practical tool to either retrieve the microscopic parameters by fitting the solutions to experiments \cite{franca2018} or to determine the requirements the materials parameter must meet in order to achieve the sought transport regime \cite{gandolfi2017emergent}.\\
\indent An abundance of analytical \cite{tzou2014macro,Ling_Li,lam2011heat,lam2013unified,tzou2010nonlocal,ramadan2009semi} and numerical works \cite{zhang2013numerical,xu2011thermal,ordonez2009thermal,torii2005heat,ramadan2009analysis} have been devoted to the DPL and CV models and their solutions in several contexts. In an effort to directly backup experimentalists, strategies to access the relevant time lags have recently been proposed, either based on specifically-designed experiments \cite{ordonez2009thermal,Kang2017} or molecular dynamics approaches \cite{Amit2015,Akbarzadeh2017}. However, the dispersion relation in $\omega$-$k$ space has remained relatively unexplored, despite the wealth of information that may be readily extracted \cite{Ordonez-Miranda2009_IJTS_2,Ordonez-Miranda2010_MRC}, namely propagating thermal wave vectors, frequencies, group velocities and quality factors.\\
\indent The present work tackles temperature propagation in the frame of the first-order DPL model, and its CV limit, adopting the same line of thought as the one that has been successfully followed in solid state physics, for instance when addressing coherent electronic transport \cite{grosso2013solid}, acoustic propagation in metamaterials \cite{Tamura1988,Giannetti2009,Travagliati2015} and electromagnetic wave propagation in matter. 
The complex-valued $\omega$-$k$ dispersion relation is investigated for the cases of a localised temperature pulse in space and of a forced temperature oscillation in time, linking the temperature wave angular frequencies $\omega$ and wave vectors $k$. A modal Q-factor is introduced to discern which temperature oscillations modes are practically accessible. The Q-factor allows mimicking the material as a frequency and/or wave-vector filter for the propagation of temperature oscillations. For the case of the temperature gradient preceding the heat flux, and in the case of the DPL (CV) model, the material acts as a bandpass (high-pass) filter for the temperature wave. The filters characteristics are accessed in terms of the relevant delay times entering the DPL model.
The Q-factors for the localized temperature pulse in space and forced temperature oscillation in time are compared and, in the former case, the group velocity is addressed.
Previous reports of temperature oscillations, among which the recently observed temperature oscillation in graphite at cryogenic temperatures \cite{huberman2019}, are revised at the light of the present formulation, together with the possibilities offered by quantum materials on the ultra-short and ultra-fast space and time scales respectively. Specifically, the possibility of observing electronic temperature wave-like behaviour in solid-state condensates and spin-temperature oscillations in magnetic materials opens the way to all-solid state thermal nanodevices, operating well above liquid helium temperature while tacking advantage of different excitations - i.e electrons, phonons and spins - and, possibly, of their mutual interplay. All the same, optimizing the conditions to observe lattice temperature wave-like oscillations in graphite further expands the range of materials amenable to new nanothermal devices schemes. The here presented approach yields to experimentalists a simple, easy-to-adopt, conceptual frame, together with algebraic formulas allowing to inspect the optimal conditions to observe temperature wave-like oscillations in materials.
The present work will be beneficial in engineering thermal devices exploiting temperature waves.\\ The overall organisation is as follows: the theoretical background is presented in section \ref{A_wave_equation_for_the_temperature}, theoretical results are discussed throughout sections \ref{Temperature_Wave_Packet_Propagation} to \ref{Comparison_between_the_two_scenarios}, case studies are addressed in section \ref{Case_studies} and conclusions summarized in section \ref{Conclusion}.

\section{\label{A_wave_equation_for_the_temperature}General dispersion relation}

\indent The starting point, to address temperature propagation, is the constitutive equation, which is obtained expanding the DPL model to first order in time:

\begin{equation}
\vet{q}\left(\vet{r},t\right)+\tau_q\frac{\partial \vet{q}}{\partial t}\left(\vet{r},t\right)=-\kappa_T\ \nabla T \left(\vet{r},t\right)-\kappa_T \tau_T\ \frac{\partial}{\partial t} \nabla T\left(\vet{r},t\right).
\label{dpl_first_order_expanded}
\end{equation}

\nd Coupling Equation \ref{dpl_first_order_expanded} with the conservation of energy\footnote{The conservation of energy is instantaneous and holds at any time $t$, no delay time hence enters the local conservation of energy.}
 at time $t$\begin{equation}
C \frac{\partial T}{\partial t}\left(\vet{r},t\right)=-\nabla \cdot \vet{q}\left(\vet{r},t\right),
\label{energy_conservation}
\end{equation}
\nd where $C$ is the volumetric heat capacity, 
and resorting to 1D propagation along the x-direction, yields the Jeffrey's type equation for the temperature scalar field \cite{tzou2014macro}:
\begin{equation}
\left( {\tau_q\over \alpha}\right)\frac{\partial^2 T}{\partial t^2}-\frac{\partial^2 T}{\partial x^2}+{1\over \alpha}\frac{\partial T}{\partial t}-\tau_T\frac{\partial^3 T}{\partial t \partial x^2}=0,
\label{temperature_wave_equation}
\end{equation}

\nd in which $\alpha=\kappa_T/C$ is the thermal diffusivity. In the present work small temperature variations, with respect to the equilibrium temperature $T_{eq}$, are addressed, hence the $\kappa_T$ and $C$ temperature dependence may be ignored. In the limit $\tau_T\rightarrow 0$  Equation \ref{temperature_wave_equation} merges into the \emph{telegraph equation}, i.e. the CV limit, while for both $\tau_T\rightarrow 0$ and $\tau_q\rightarrow 0$ the classical Fourier diffusion equation is restored.\\
\indent Adimensionalization of Equation \ref{temperature_wave_equation} is conveniently achieved by introducing the set of non-dimensional variables
$\beta=\frac{t}{\tau_q}$, $\xi=\frac{x}{\sqrt{\alpha \tau_q}}$, $Z=\frac{\tau_T}{\tau_q}$, $\theta=\frac{T}{T_{eq}}$, hence yielding:
\begin{equation}
\frac{\partial^2 \theta}{\partial \beta^2}-\frac{\partial^2 \theta}{\partial \xi^2}+\frac{\partial \theta}{\partial \beta}-Z\frac{\partial^3 \theta}{\partial \beta \partial \xi^2}=0,
\label{temperature_wave_equation_adimension}
\end{equation}

\nd We seek for solutions of Equation \ref{temperature_wave_equation_adimension} in the form $\theta\left(\xi,\beta\right)=\theta_0e^{i(\tilde{k}\xi+\tilde{\omega} \beta)}$, where the complex-valued adimensional wave vector, $\tilde{k}$, and angular frequency, $\tilde{\omega}$, are linked to their dimensional counterparts $k$ and $\omega$ by: $\tilde{\omega}=\tau_{q}\omega$ and $\tilde{k}=\left(\sqrt{\alpha \tau_q}\right)k$.
Substituting $\theta\left(\xi,\beta\right)$ in Equation \ref{temperature_wave_equation_adimension} gives the dispersion relation:

\begin{equation}
\tilde{k}^2\left(1+iZ\tilde{\omega} \right)=\tilde{\omega}^2\left(1-\frac{i}{\tilde{\omega}} \right).
\label{complex_dispersion}
\end{equation}

Already in its general form, the dispersion relation provides, in a straight-forward manner, the general conditions \cite{tang1999wavy} that must be met in order to observe wave-like temperature propagation.
For $Z|\tilde{\omega}| \ll 1$ and $1/|\tilde{\omega}| \ll 1$, Equation \ref{complex_dispersion} reduces to the dispersion relation $\tilde{k}^2=\tilde{\omega}^2$ for a free-propagating wave.
Rearranging terms, the above mentioned prescription may be cast as $Z \ll 1/|\tilde{\omega}| \ll 1$.
Switching back to dimensional variables, the condition for wave-like motion reads $\tau_{T} \ll 1/|\omega| \ll \tau_{q}$, meaning that the temperature oscillation period in time $2\pi/|\omega|$ must lay between the two relaxation times and  that the temperature gradient must precede the onset of heat flux.

The dispersion relation in a linear problem, as in the present case, does not depend on the spatio-temporal features of the excitation. Nevertheless, the dispersion relation in its general form is too involved to allow making further progress. Depending on the excitation scenario though, some features of the dispersion relation may be simplified, allowing for further understanding of the propagation problem.

\begin{figure}[H]
\begin{center}
\includegraphics[width=0.6\textwidth]{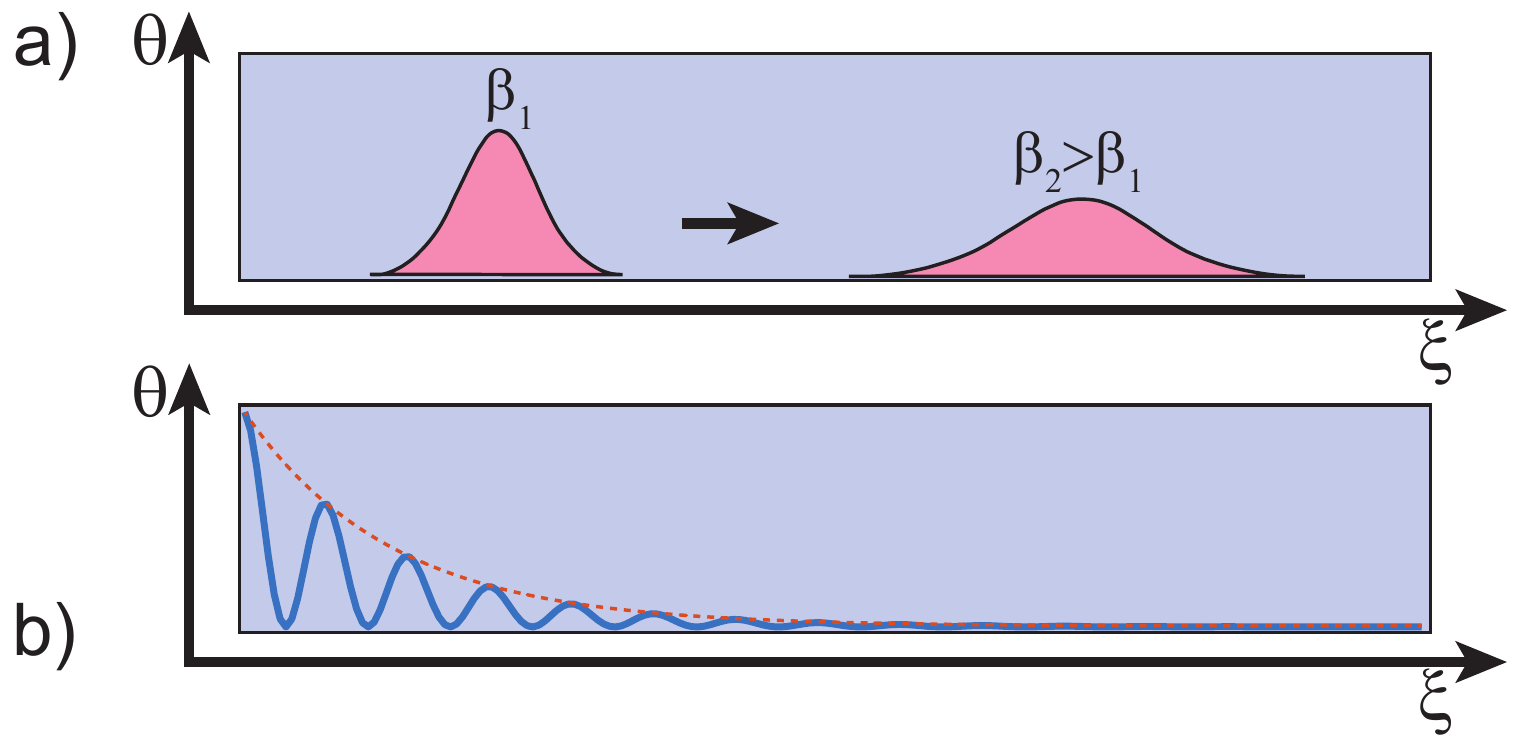} 
\caption{Schematics of (a) temperature pulse in space and (b) forced temperature oscillation in time.}
\label{pulses}
\end{center}
\end{figure}

In the present paper we address the propagation of a spatial temperature peak and a forced temperature oscillation in time.\\
A temperature pulse in space (Figure \ref{pulses} a) may be obtained as a linear superposition of infinitely many spatial-harmonic oscillations, each of the form $e^{i\tilde{k}\xi}$, with the non-dimensional wave-vector $\tilde{k}$ being the integrating variable in the integral.
It thus suffices to investigate the time-evolution of each space-harmonic component. As time evolves, each space-harmonic component remains periodic in space while, due to damping, its amplitude is reduced in time. This scenario is accounted for assuming a complex-valued angular frequency $\tilde{\omega}$ and a real-valued wave-vector $\tilde{k}$. This case is addressed in Section \ref{Temperature_Wave_Packet_Propagation}.\\

\indent A forced temperature oscillation in time (Figure  \ref{pulses} b) may be obtained by forcing the temperature at a specific location to oscillate at an angular frequency $\tilde{\omega}$, for instance by forcing the temperature to oscillate at the surface of a semi-infinite solid. Moving away from the excitation source, the temperature still oscillates at the same temporal frequency, whereas the amplitude, due to damping, diminishes with increasing distance from the source. This scenario is accounted for by assuming a complex-valued wave-vector $\tilde{k}$ and a real-valued angular frequency $\tilde{\omega}$. This case is addressed in Section \ref{Temperature_Sustained_Wave-propagation}. The two scenarios are compared in Section \ref{Comparison_between_the_two_scenarios}.\\

\section{\label{Temperature_Wave_Packet_Propagation}Dispersion relation: spatial temperature pulse}

\indent In the following, we investigate the propagation of a spatial temperature pulse. The angular frequency is denoted $\tilde{\omega}=\tilde{\omega}_1+i\tilde{\omega}_2$, where $|\tilde{\omega}_1|=2\pi/\beta_{osc}$ is the inverse of the oscillation period $\beta_{osc}$ and $\tilde{\omega}_2=1/\beta_{damp}$ is the inverse of the damping time $\beta_{damp}$, the latter giving a measure of the maximum time-lapse within which temperature oscillations can be observed before the propagation turns diffusive; $\tilde{k}\in\mathbb{R}$.\\
\indent  The onset of a wave-like regime is conveniently addressed introducing the Q-factor, $Q=|\omega_1|/\omega_2=|\tilde{\omega}_1|/\tilde{\omega}_2$, which discriminates the underdamped ($Q>1$) from the overdamped ($0<Q<1$) and the non-oscillatory regime ($Q=0$).\\
In the following, the expression for $\tilde{\omega}(\tilde{k})$, $Q(\tilde{k})$ 
are first derived for the case of $Z=0$ (CV limit) and next for the general case $Z\ne0$. This allows illustrating the effects introduced by the additional delay time $\tau_{T}$ with respect to the CV case.

\subsection{CV Model Limit \label{CVmodelLimit_packet}} 

\indent Substituting $\tilde{\omega}=\tilde{\omega}_1+i\tilde{\omega}_2$ into the dispersion relation Equation \eqref{complex_dispersion}, calculated for $Z=0$, stems in:\\

\begin{equation}
\left\{\begin{array}{l}
\tilde{\omega}_1^2-\tilde{\omega}_2^2+\tilde{\omega}_2-\tilde{k}^2=0\\
\\
\left(2\tilde{\omega}_2-1\right)\tilde{\omega}_1=0.\\
\end{array}\right.
\label{real0imag0}
\end{equation}

We first seek for potentially oscillatory solutions, i.e. $\tilde{\omega}_1\neq0$. The second equation of the above system gives $\tilde{\omega}_2=1/2$, and, upon substitution of $\tilde{\omega}_2=1/2$, the first equation of the system reduces to:
\begin{equation}
\ \tilde{\omega}_1^2-\tilde{k}^2+\frac{1}{4}=0.
\label{dispersion_for_omega1}
\end{equation}
\indent Equation \ref{dispersion_for_omega1} admits real and non-null solutions provided the discriminant $\Delta(\tilde{k})=\tilde{k}^2-\frac{1}{4}>0$.
The latter condition is satisfied for wave vectors in the range $|\tilde{k}|>\tilde{k}_{lo}=1/2$.
The potentially oscillatory solutions and their Q-factors are: 

\begin{equation}
\left\{\begin{array}{ll}
\tilde{\omega}_1(\tilde{k}) =\pm\sqrt{\Delta(\tilde{k})}=\pm\sqrt{\tilde{k}^2-\displaystyle{\frac{1}{4}}}&\\
\displaystyle{\tilde{\omega}_2(\tilde{k})=\displaystyle{\frac{1}{2}}},&\\
\end{array}\right.
\label{disp_tauTlim0CV}
\end{equation}

\begin{equation}
Q(\tilde{k})=\sqrt{4\tilde{k}^2-1.\ } 
\label{Q_factTlim0CV}
\end{equation}

\begin{figure*}
\begin{center}
\includegraphics[width=1\textwidth]{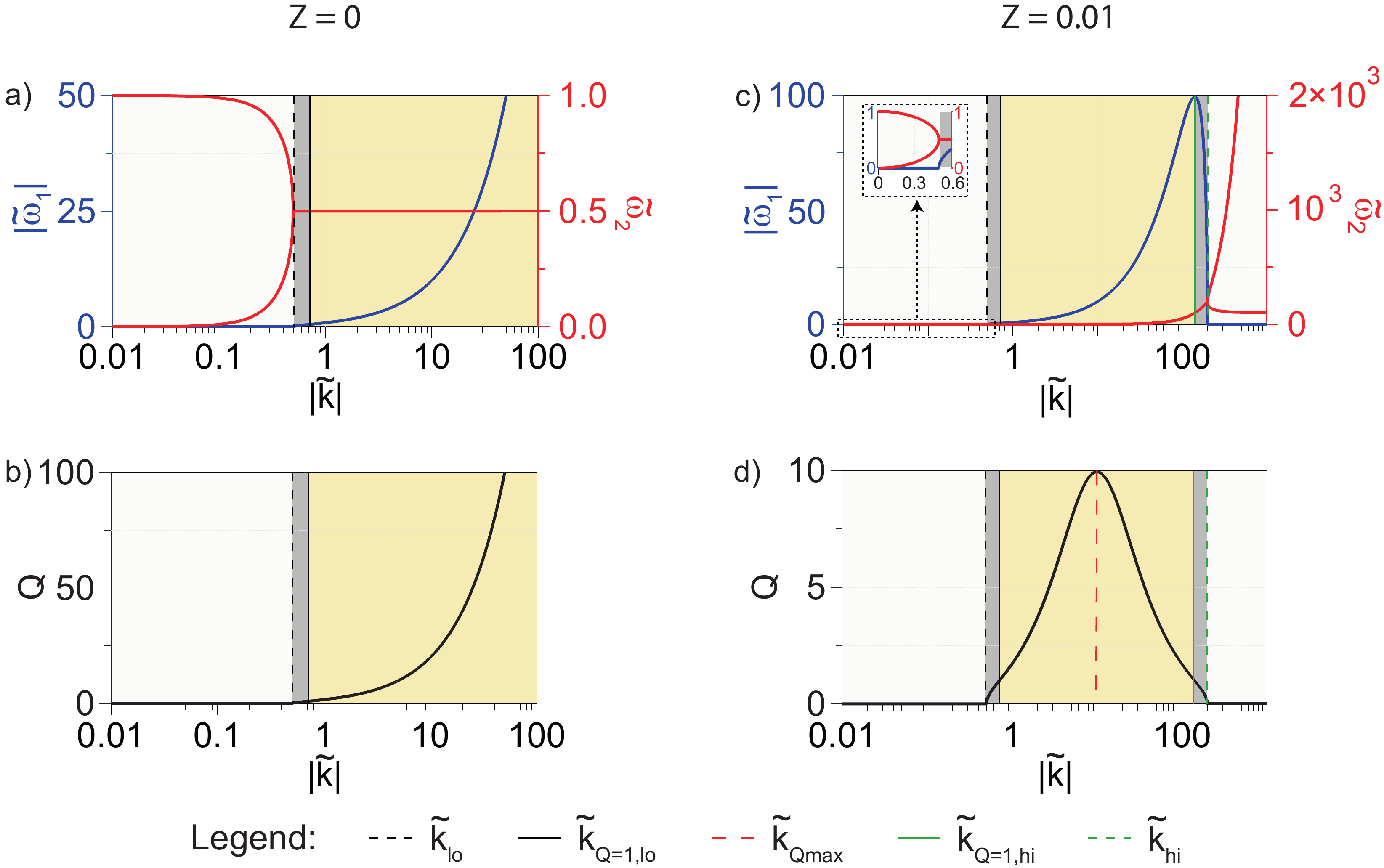} 
\caption{Dispersion relation and Q-factor for the $\tilde{\omega}\in\mathbb{C}$ and $\tilde{k}\in\mathbb{R}$ case. Lin-log scale plots are adopted unless otherwise stated. (a) plot of the dispersion relation: $|\tilde{\omega}_1|$ (blue line, left axis) and $\tilde{\omega}_2$ (red line, right axis) vs $|\tilde{k}|$ (horizontal axis, log scale) for the case of $Z=0$, i.e. the CV model. (b) plot of the Q-factor vs $|\tilde{k}|$ for the case of $Z=0$, i.e. the CV model. 
(c) and (d): same as for panels (a) and (b) respectively but for the case $Z=0.01$. The inset of panel (c) represents the expanded view of $\tilde{\omega}_1$ and  $\tilde{\omega}_2$ for low wave vectors in lin-lin scale (as opposed to the lin-log scale of the main graph).
The regions admitting underdamped (i.e. $Q>1$), overdamped (i.e. with $0<Q<1$) and non-oscillatory solutions are highlighted in yellow, dark-grey and left blank respectively.
The black dashed (continuous) line denotes $\tilde{k}_{lo}$ ($\tilde{k}_{Q=1,lo}$), the green dashed (continuous) line indicates $\tilde{k}_{hi}$ ($\tilde{k}_{Q=1,hi}$). The red dashed line in panel (d) denotes the wave vector $\tilde{k}_{Qmax}$ for which the $Q$-factor is maximum.}
\label{complex_omega_dispersion}
\end{center}
\end{figure*}

For the case in which no oscillatory solutions are admitted, i.e $\tilde{\omega}_{1}=0$, the first equation in system (\ref{real0imag0})
reduces to
\begin{equation}
\tilde{\omega}_2^2-\tilde{\omega}_2+\tilde{k}^2=0,
\label{equation_omega2}
\end{equation}
the latter relation allowing real solutions for $\tilde{\omega}_2$ provided $\Delta(\tilde{k})\leq0$ (we pinpoint that the discriminant of Equation \ref{equation_omega2} is equal to $-\Delta(\tilde{k})$).
The latter condition is satisfied for wave vectors in the range $|\tilde{k}|\le\tilde{k}_{lo}=1/2$.
Thus, the non-oscillatory solutions are:\\
\begin{equation}
\left\{\begin{array}{ll}
\tilde{\omega}_1(\tilde{k}) =0&\\
& \\
\displaystyle{\tilde{\omega}_2(\tilde{k})}=\displaystyle{\frac{1}{2}\pm\sqrt{-\Delta(\tilde{k})}}=\displaystyle{\frac{1}{2}\pm\sqrt{\frac{1}{4}-\tilde{k}^2}},&\\
\end{array}\right.
\label{disp_tauTlim0CV_imag}
\end{equation}
whereas the Q-factor is null. 
Throughout the manuscript the subscripts $lo$ (for low) and $hi$ (for high) will define the lower and higher boundary value of a given range that will be specified case by case. For the case of the CV model only the subscript $lo$ will appear, since the upper range will diverge to infinity in all considered cases.

The dispersion relation is reported in lin-log scale in Figure \ref{complex_omega_dispersion} a where the oscillation angular frequency $|\tilde{\omega}_1|$ (blue curve, left axis) and the inverse damping time $\tilde{\omega}_2$ (red curve, right axis) are plotted vs the wave vector $|\tilde{k}|$.
As for the use of the absolute values in the plots, we pinpoint that for a given $\tilde{k}$ value there are actually two opposite values of $\tilde{\omega}_{1}$, whereas for a given $\tilde{\omega}_{1}$ one finds two opposite values of $\tilde{k}$, see SI for further details. 
For $|\tilde{k}|>\tilde{k}_{lo}$ different modes have different oscillation angular frequencies but same damping time, whereas for $|\tilde{k}|\leq\tilde{k}_{lo}$ two non-oscillatory solutions are admitted.\\
\indent The meaning of these solutions, with regard to the wave-like behavior, is better synthetised in Figure  \ref{complex_omega_dispersion} b, where the Q-factor is plotted vs $|\tilde{k}|$ in lin-log scale. The plot shows that the material behaves as a \textit{high-pass filter} for temperature oscillations. Specifically, for $|\tilde{k}|>\tilde{k}_{Q=1,lo}=1/\sqrt{2}$ one has $Q>1$, that is the material sustains underdamped wave-like temperature oscillations (yellow shaded portion). 
Furthermore, $Q\sim2|\tilde{k}|$ for $|\tilde{k}|\gg1/\sqrt{2}$ (mind the fact that a lin-log scale is adopted in Figure \ref{complex_omega_dispersion}), thus approaching the case of free-wave propagation: $| \tilde{\omega}_1|\simeq| \tilde{\omega}|\sim|\tilde{k}|$.
For $\tilde{k}_{lo}<|\tilde{k}|<\tilde{k}_{Q=1,lo}$ (dark gray shaded portion) one has $0<Q<1$, that is temperature oscillations are overdamped, no wave-like behavior is therefore experimentally accessible. For $|\tilde{k}|\leq\tilde{k}_{lo}$ (white portion) one has $Q=0$, i.e. $\tilde{\omega}_{1}=0$.
From the perspective of observing temperature wave-like propagation the latter two cases are substantially alike. The high-pass filter characteristic holds also with respect to $\tilde{\omega}_{1}$, as can be seen inspecting the relation between $|\tilde{\omega}_{1}|$ and $|\tilde{k}|$, the blue curve in Figure  \ref{complex_omega_dispersion} a.\\
\indent Resorting to dimensional variables, the condition $|\tilde{k}|\leq\tilde{k}_{lo}=1/2$ has the vivid physical meaning $\lambda\geq4\pi D_{q}$, where $\lambda$=2$\pi /k$ is the thermal wavelength and $D_{q}=\sqrt{\alpha\tau_{q}}$ the diffusion length.
Where the thermal wavelength exceeds the diffusion length by a factor of $4\pi$, the oscillatory behaviour is suppressed.
Furthermore, for $|\tilde{k}|\gg 1/2$ on has $\lambda\ll D_{q}$ and $Q\sim D_{q}/\lambda$, the meaning being that, when the thermal wavelength is very short with respect to the diffusion length, the solution becomes purely oscillatory. A transition region exists in between the two regimes, where, although $\lambda$ is small enough to be out of the diffusive regime, $\lambda<4\pi D_{q}$, it is still too long to develop underdamped oscillations, $\lambda>2\sqrt{2}\pi D_{q}$.\\

\subsection{General case\label{subsTauTminTauq_packetZ}}

\indent In the following, the investigation is extended to the general case $Z\neq0$, proceeding in analogy with the derivation reported in the previous section. The complex eigenfrequency $\tilde{\omega}=\tilde{\omega}_1+i\tilde{\omega}_2$ is substituted into dispersion equation \eqref{complex_dispersion}.\\
\indent We first seek for potentially oscillatory solutions, i.e. $\tilde{\omega}_1\neq0$. A second degree algebraic equation for $\tilde{\omega}_1$, characterised by the discriminant $\Delta(\tilde{k})$, is thus obtained (see SI for details):
\begin{equation}
\Delta(\tilde{k})=-\frac{Z^2}{4} \tilde{k}^4+\left(1-{Z\over2} \right)\tilde{k}^2-\frac{1}{4}.
\label{delta}
\end{equation}
\indent In order for $\tilde\omega_{1}$ to be real and non-null, the inequality $\Delta(\tilde{k})>0$ must hold. The latter condition can be fulfilled for a specific set of  wave vectors if and only if $Z<1$, i.e. if the temperature gradient precedes the heat flux. Specifically, $\Delta(\tilde{k})$ is positive only for $|\tilde{k}|$ in the range $\tilde{k}_{lo}<|\tilde{k}|<\tilde{k}_{hi}$, where:
\begin{equation}
\tilde{k}_{lo(hi)}=\sqrt{\frac{2}{Z^2}\left[1-\frac{Z}{2}-(+)\sqrt{1-Z}\ \right]}.\\
\label{klow}
\end{equation}
\nd Therefore, differently from the CV limit, the general case introduces a second cutoff, $\tilde{k}_{hi}$, for the wave vectors beyond which no oscillatory solution is admitted. The potentially oscillatory solutions and their Q-factor read: 

\begin{equation}
\left\{\begin{array}{l}
\displaystyle{\tilde{\omega}_1 =\pm\sqrt{\Delta(\tilde{k})}}=\pm\sqrt{-\frac{Z^2}{4} \tilde{k}^4+\left[1-{Z\over2} \right]\tilde{k}^2-\frac{1}{4}}\\
\\
\displaystyle{\tilde{\omega}_2=\frac{1}{2}+\frac{Z}{2}\tilde{k}^2},\\
\end{array}
\right.
\label{disp_tauTmintauq_DPL}
\end{equation}
\begin{equation}
Q(\tilde{k})=\sqrt{\frac{4 \tilde{k}^2}{\left(1+Z \tilde{k}^2 \right)^2}-1}.
\label{Q_factor_TauqmagTauT_DPL}
\end{equation}

\indent No oscillatory solution is admitted, i.e $\tilde{\omega}_{1}=0$, provided $\Delta(\tilde{k})\leq0$. 
If $Z\geq 1$, the latter condition is fulfilled for every wave vector.
On the other hand, if $Z<1$, we have $\Delta(\tilde{k})\leq0$ only for wave vector in the range $|\tilde{k}|\leq\tilde{k}_{lo}$ and $|\tilde{k}|\geq\tilde{k}_{hi}$. 
The dispersion relation for non-oscillatory modes reads:

\begin{equation}
\left\{\begin{array}{l}
\displaystyle{\tilde{\omega}_1 =0}\\
\\
\displaystyle{\tilde{\omega}_2=\frac{1}{2}+\frac{Z}{2}\tilde{k}^2\pm\sqrt{-\Delta(\tilde{k})}},\\
\end{array}
\right.
\label{disp_tauTmintauq_DPL_non_oscillating}
\end{equation}

\noindent whereas the Q-factor is zero.\\
\begin{figure*}[t]
\begin{center}
\includegraphics[width=1\textwidth]{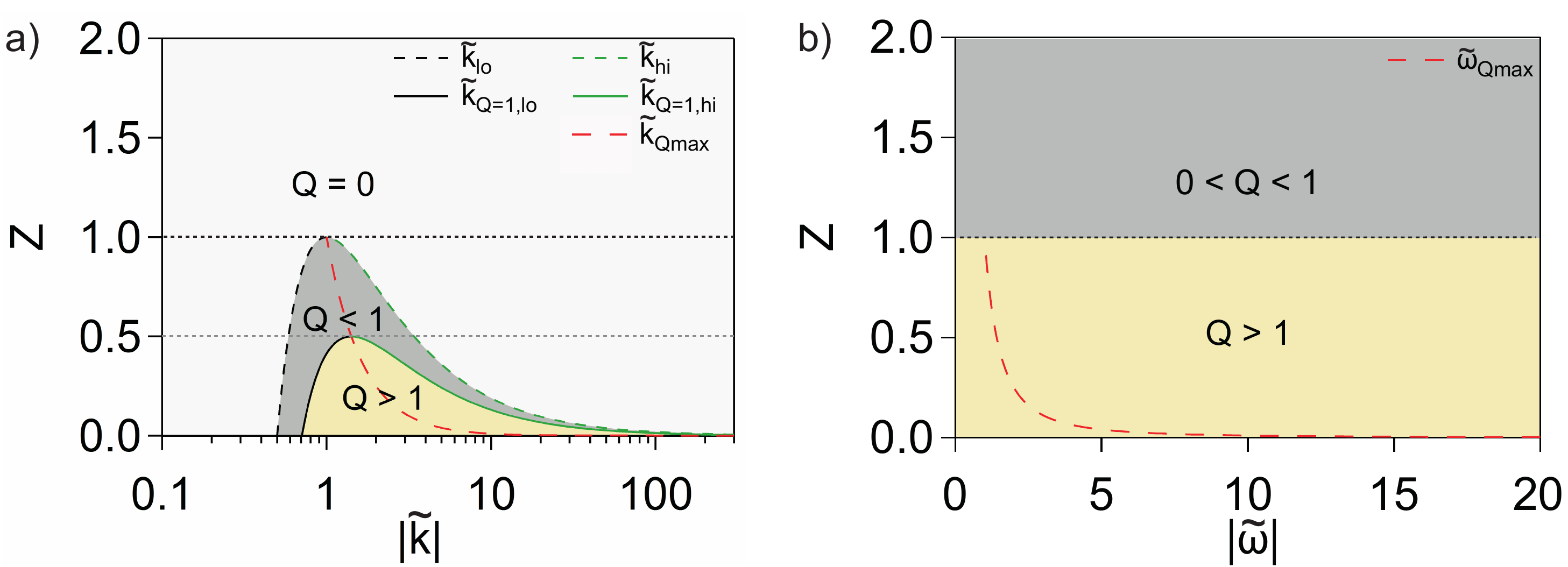} 
\caption{Partitioning of the the $\tilde{k}-Z$ plane (panel a) and $\tilde{\omega}-Z$ plane (panel b) in regions were the temperature oscillation is underdamped (yellow), overdamped (dark gray) and non-oscillating (white). (a) $\tilde{\omega}\in\mathbb{C}$ and $\tilde{k}\in\mathbb{R}$ case. A lin-log scale is adopted. 
The red-dashed line represents the curve $\tilde{k}_{Qmax}$ vs $Z$, the black-dashed (continuous) line represents $\tilde{k}_{lo}$ ($\tilde{k}_{Q=1,lo}$) vs $Z$, the green-dashed (continuous) line represents $\tilde{k}_{hi}$ ($\tilde{k}_{Q=1,hi}$) vs $Z$.
(b) $\tilde{k}\in\mathbb{C}$ and $\tilde{\omega}\in\mathbb{R}$ case. A lin-lin scale is adopted. The red-dashed line represents the curve $\tilde{\omega}_{Qmax}$ vs $Z$.}
\label{Q_zero_or_one}
\end{center}
\end{figure*}
\indent The dispersion relation and Q-factor are functions of both $\tilde{k}$ and $Z$, see SI for the plot of $Q(\tilde{k},Z)$.
Since we are interested in observing wave-like temperature oscillations, we discriminate among the possible cases inspecting the regions in $\tilde{k}-Z$ space in which $Q>1$, $0<Q<1$ and $Q=0$. 
The solutions of the above-mentioned inequalities are reported in Figure \ref{Q_zero_or_one} a, where the values in the $\tilde{k}-Z$ plane admitting underdamped (yellow), overdamped (dark gray) and non-oscillatory (white) modes are shown.

For $0\le Z<1/2$ there always exists a range of wave vectors $\tilde{k}$ for which \textit{underdamped} wave-like temperature oscillations are admitted. For $1/2\leq Z<1$ underdamped wave-like temperature oscillations are suppressed, nevertheless there still exists a range of wave vectors for which \textit{overdamped} wave-like temperature oscillations are admitted. For $Z\geq1$ no \textit{oscillating} solutions whatsoever are admitted. For a finite value of $Z$, the Q-factor is now bound and its maximum value is obtained for $\tilde{k}=\pm \tilde{k}_{Qmax}=\pm\left(Z\right)^{-1/2} $, yielding:
\begin{equation}
Q\left(\pm\  \tilde{k}_{Qmax}\right)=\sqrt{{1\over Z}-1}.
\label{Qmax}
\end{equation}

\noindent In Figure \ref{Q_zero_or_one} a, the curve $\tilde{k}_{Qmax}$ is plotted as a function of $Z$, see red dashed line in Figure \ref{Q_zero_or_one} a.
The CV limit is restored for $Z\rightarrow 0$ where $\tilde{k}_{hi}\rightarrow +\infty$ and $\tilde{k}_{Qmax}\rightarrow +\infty$, and the maximum for the Q-factor diverges.\\

\indent We now focus on the case in which under-damped wave-like temperature oscillations may be admitted, i.e. $0<Z<1/2$. The dispersion relation - $|\tilde{\omega}_{1}|$ (blue curve, left axis) and $\tilde{\omega}_{2}$ (red curve, right axis) - and the Q-factor are plotted vs $|\tilde{k}|$ in lin-log scale in Figure \ref{complex_omega_dispersion} c and d respectively, where a value of $Z=0.01$ has been chosen for the sake of illustrating the salient features.\\
\indent As for the dispersion, the striking difference, brought in by the non-zero delay time $\tau_{T}$, is the onset of an upper cutoff wavelength, $\tilde{k}_{hi}$, beyond which no-oscillatory behaviour is admitted, see Figure \ref{complex_omega_dispersion} c. 
Furthermore, the low-frequency cut-off, $\tilde{k}_{lo}$, is now $Z$-dependent.
The maximum oscillation frequency is also bound and occurs at a wave vector $\tilde{k}=\pm \tilde{k}_{max}$ where:
\begin{equation}
\tilde{k}_{max}=\sqrt{\frac{1}{Z}\left(\frac{2}{Z}-1\right)}.
\label{kmax}
\end{equation}

\indent As for the $Q$-factor, it shows that the material behaves as a \textit{bandpass} filter for temperature oscillations, a fact well illustrated in Figure \ref{complex_omega_dispersion} d. 
The $Q$-factor reported in Figure \ref{complex_omega_dispersion} b and d corresponds to the cut taken along the line $Z=0$ and $Z=0.01$ respecrively in Figure \ref{Q_zero_or_one} a.\\
\indent Let's analyse the filter characteristics for the general case $0<Z<1/2$. Underdamped wave-like temperature oscillations may be observed provided $Q(\tilde{k})>1$ (yellow shaded region in Figure \ref{complex_omega_dispersion} d for $Z=0.01$). The latter is satisfied if $Z<1/2$ and $\tilde{k}_{Q=1,lo}<|\tilde{k}|<\tilde{k}_{Q=1,hi}$ (see SI for the solution of the  inequality), where:
\begin{equation}
\tilde{k}_{Q=1,lo (hi)}= \frac{\sqrt{(1-Z)-(+)\sqrt{1-2Z}}}{Z},
\label{Q=1}
\end{equation}
thus defining the filter pass-band $[\tilde{k}_{Q=1,lo},\tilde{k}_{Q=1,hi}]$.
The third-order term in Equation \ref{temperature_wave_equation}, brought in by the delay time $\tau_{T}$, thus hampers temperature oscillations both on the  low and high wave vectors side.
On the low side, $\tilde{k}_{Q=1,lo}$ ranges from $1/\sqrt{2}$ for Z=0 (i.e CV case: $\tau_{T}$= 0) to $\sqrt{2}$ for Z=1/2. The effect is drastic on the high wave vector side, where $\tilde{k}_{Q=1,hi}$, which diverges for $Z\rightarrow0$, approaches $\sqrt{2}$ for Z=1/2. The bandpass filter characteristic holds also with respect to $\tilde{\omega}_{1}$, as can be appreciated mapping $\tilde{\omega_{1}}$ vs $\tilde{k}$, see blue curve in Figure \ref{complex_omega_dispersion} c, and plotting the $Q$-factor accordingly.\\
\indent The dispersion relation allows addressing the group velocity for the temperature pulse. The definition of an adimensional group velocity, $\tilde{v}_g(\tilde{k})=\partial \tilde{\omega}_1/\partial \tilde{k}$, remains physically meaningful provided the temperature pulse distortion during propagation is not too severe. For instance, if damping selectively suppresses Fourier components at certain $\tilde{k}$ vectors with respect to others, the initial pulse's center-of-mass looses significance all together with the concept of group velocity. The concept of group velocity thus remains valid provided $Q\gg1$.
The sign of $\tilde{v}_g$ changes in correspondence of the $\tilde{k}$ values maximising $\tilde{\omega}_{1}$, $|\tilde{k}|=\tilde{k}_{max}$, a fact well appreciated inspecting the plot of $|\tilde{\omega}_1|$ vs $|\tilde{k}|$ reported in Figure \ref{complex_omega_dispersion} c or upon direct inspection of its expression:
\begin{equation}
\tilde{v}_g(\tilde{k})=\frac{\partial \tilde{\omega}_1}{\partial \tilde{k}}=\pm \frac{1}{2}\frac{-Z^2 \tilde{k}^3+2\left[1-{Z\over2} \right]\tilde{k}}{\sqrt{-\frac{Z^2}{4} \tilde{k}^4+\left[1-{Z\over2} \right]\tilde{k}^2-\frac{1}{4}}}.
\label{group_velocity_Znon0}
\end{equation}
For the sake of simplicity let's focus on the band's branch characterised by $\tilde{k}\ge0$ and $\tilde{\omega}\ge0$. The group velocity is positive for $\tilde{k}<\tilde{k}_{max}$ and negative for $\tilde{k}>\tilde{k}_{max}$. The latter $\tilde{k}$ range falls in the overdamped region where the wave-packet is suppressed and $\tilde{v}_{g}$ loses its significance. This fact shows that, within a given branch and for practical purposes, the group velocity preserves the same sign.

\section{\label{Temperature_Sustained_Wave-propagation}Dispersion relation: Forced Temperature Oscillation in Time}
\begin{figure*}[t]
\begin{center}
\includegraphics[width=1\textwidth]{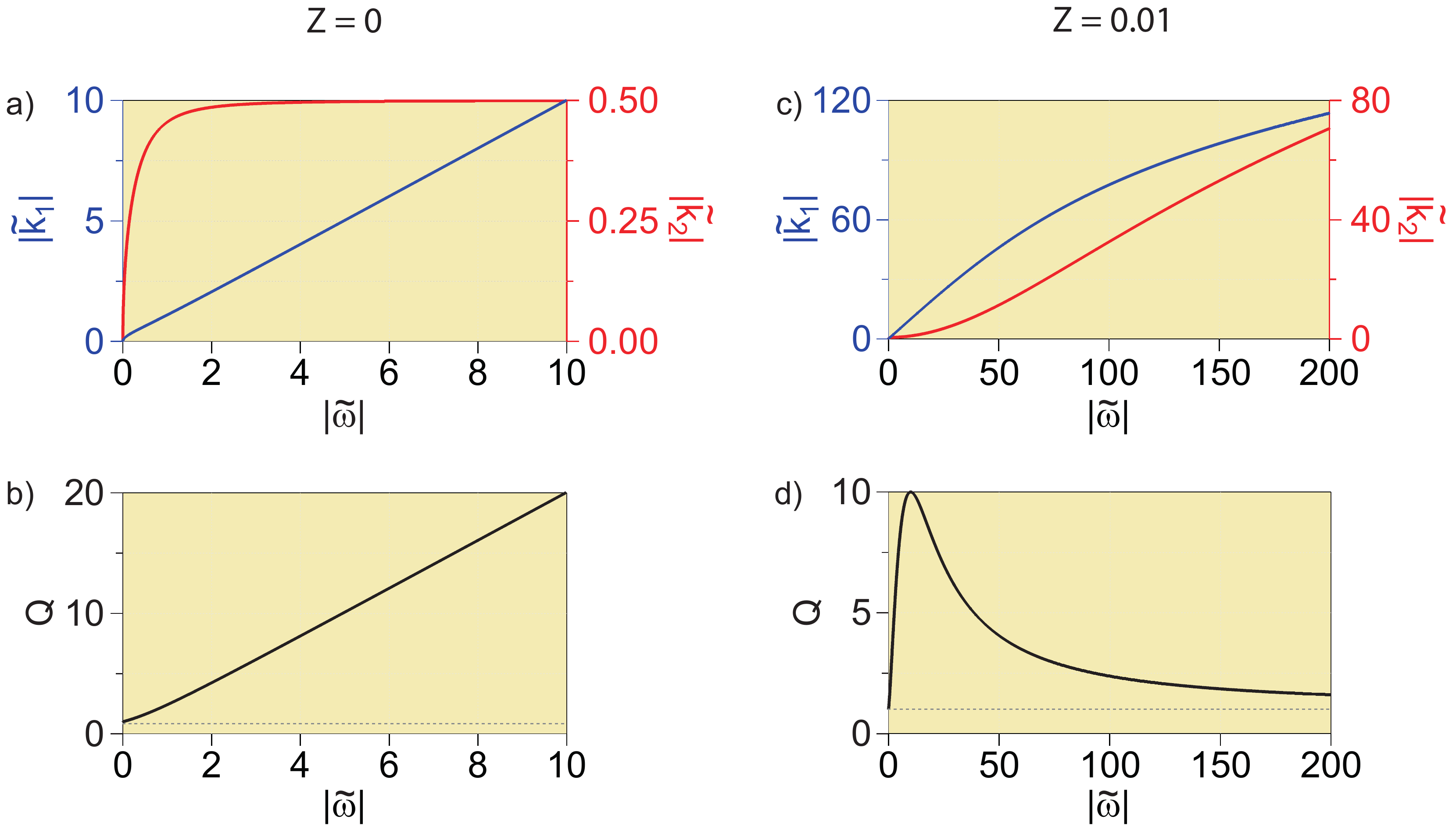} 
\caption{Dispersion relation and Q-factor for the case $\tilde{k}\in\mathbb{C}$ and $\tilde{\omega}\in\mathbb{R}$. Lin-lin scale plots are adopted. (a) dispersion relation $|\tilde{k}_1|$ (blue line, left axis) and $|\tilde{k}_2|$ (red line, right axis) vs $|\tilde{\omega}|$ (horizontal axis) for $Z=0$, i.e CV model. (b) Q-factor vs $|\tilde{\omega}|$; the $Q=1$ reference line (dotted grey line) indicates underdamped oscillatory solutions (yellow shading of the panels consistently with Figure \ref{complex_omega_dispersion}). (c) and (d): same as for panel (a) and (b) respectively but for the case $Z=0.01$.}
\label{complex_k_dispersion}
\end{center}
\end{figure*}
\noindent In this section we investigate a forced temperature oscillation in time.
The wave vector is denoted $\tilde{k}=\tilde{k}_1+i\tilde{k}_2$, where $|\tilde{k}_1|=2\pi/\xi_{osc}$ is the inverse of the oscillation length $\xi_{osc}$ and $\tilde{k}_2=1/\xi_{damp}$ is the inverse of the damping length $\xi_{damp}$,  the latter giving a measure of the maximum distance from the excitation point at which the temperature oscillations can be observed;
 $\tilde{\omega}\in\mathbb{R}$.\\
\indent  We introduce the Q-factor, defined in the current case as $Q=|k_1|/|k_2|=|\tilde{k}_1|/|\tilde{k}_2|$, discriminating the underdamped ($Q>1$) from the overdamped ($0<Q<1$) and the non-oscillatory regime ($Q=0$).\\
We derive the expression for $\tilde{k}(\tilde{\omega})$, $Q(\tilde{\omega})$ first in the case of $Z=0$ (CV limit) and afterwards for the general case $Z\ne0$.

\subsection{CV Model Limit \label{CVmodelLimit_forced}} 
\indent Proceeding in analogy with the derivation reported in Section \ref{CVmodelLimit_packet}, we substitute $\tilde{k}=\tilde{k}_1+i\tilde{k}_2$ into the dispersion relation Equation \eqref{complex_dispersion}, calculated for $Z=0$.\\
Performing the algebra (see SI for further details) one proves that, for every non-zero angular frequency $\tilde{\omega}$, all the modes have a non-zero oscillatory wave vector $\tilde{k}_1$, that is, at variance with the temperature spatial pulse case, no frequency cut-offs are present.
The solutions and the Q-factor read:

\begin{equation}
\left\{\begin{array}{l}
\displaystyle{\tilde{k}_1 =\mp \sqrt{\frac{\left|\tilde{\omega}\right|}{2}\sqrt{1+\tilde{\omega}^2}+\frac{\tilde{\omega}^2}{2}}}\\
\\
\displaystyle{\tilde{k}_2=\pm sign(\tilde{\omega})\sqrt{\frac{\left|\tilde{\omega}\right|}{2}\sqrt{1+\tilde{\omega}^2}-\frac{\tilde{\omega}^2}{2}}},\\
\end{array} \right.
\label{disp_kcomplex_CV}
\end{equation}
\begin{equation}
Q(\tilde{\omega})=\left|\tilde{\omega}\right|+\sqrt{\tilde{\omega}^2+1}.
\label{qualtifactor_k_complex_CV}
\end{equation}

The dispersion relation is reported in Figure \ref{complex_k_dispersion}, where the absolute value of the oscillations wave vector $|\tilde{k}_1|$ (blue curve, left axis) and that of the inverse damping length $|\tilde{k}_2|$ (red curve, right axis) are plotted against the absolute value of the angular frequency $|\tilde{\omega}|$. In plotting the graphs, at variance with the temperature pulse case, a lin-lin scale is here adopted for ease of visualisation, whereas the absolute values are used for the same reasons as in the preceding case\footnote{As we did for the spatial temperature pulse, we do not keep into account $\tilde{\omega}$, $\tilde{k}_1$ and $\tilde{k}_2$ signs, the latter discriminating between modes with the same properties but propagating in two different directions (namely forward or backward propagating).}. 

\indent The meaning of these solutions, with regard to the wave-like behavior, is shown in Figure  \ref{complex_k_dispersion} b, where the Q-factor is plotted vs $|\tilde{\omega}|$: $Q\geq1$ for every angular frequency and increases with $|\tilde{\omega}|$. For $|\tilde{\omega}|\gg1$ one has $Q\sim2|\tilde{\omega}|$, thus approaching the case of free-wave propagation:
$|\tilde{k}_1|\simeq| \tilde{k}|\sim|\tilde{\omega}|$.
Although, formally speaking, in the present case the material always sustains underdamped oscillations, from an experimental stand point one is better off with $Q\gg1$. In this sense the system behaves as an actual \textit{high-pass filter} for temperature oscillation frequencies, as displayed in Figure \ref{complex_k_dispersion} b. The high-pass filter characteristic holds also with respect to $|\tilde{k}_{1}|$.

\subsection{General case \label{subsTauTminTauq_sustained}}

\indent Also for the general case $Z\neq0$, all modes have a non-zero oscillatory wave vector $\tilde{k}_1$ (refer to SI for the derivation). The dispersion relation and Q-factor read: 

\begin{equation}
\left\{\begin{array}{l}
\displaystyle{\tilde{k}_1 =\mp \sqrt{\frac{\left|\tilde{\omega}\right|}{2}\sqrt{\frac{1+\tilde{\omega}^2}{1+Z^2\tilde{\omega}^2}}-\frac{\tilde{\omega}^2}{2}\left(\frac{Z-1}{1+Z^2\tilde{\omega}^2}\right)}\ \ }\\
\\
\displaystyle{\tilde{k}_2=\pm sign(\tilde{\omega})\sqrt{\frac{\left|\tilde{\omega}\right|}{2}\sqrt{\frac{1+\tilde{\omega}^2}{1+Z^2\tilde{\omega}^2}}+\frac{\tilde{\omega}^2}{2}\left(\frac{Z-1}{1+Z^2\tilde{\omega}^2}\right)},}\\
\end{array} \right.
\label{disp_kcomplex_omegareal_norm}
\end{equation}
\begin{equation}
Q(\tilde{\omega})=\frac{\left|\tilde{\omega}\right|\left(1-Z\right)}{1+Z\tilde{\omega}^2}+\sqrt{\left[\frac{\left|\tilde{\omega}\right|\left(1-Z\right)}{1+Z\tilde{\omega}^2}\right]^2+1}.
\label{qualtifactor_dependent_on_omega_norm}
\end{equation}

\indent The dispersion relation and Q-factor being functions of both $\tilde{\omega}$ and $Z$, the kind of oscillatory solution is discriminated partitioning the $\tilde{\omega}-Z$ space according to the value of $Q$, see Figure \ref{Q_zero_or_one} b (refer to SI for the full plot of $Q(\tilde{\omega},Z)$).

For $0\le Z<1$ one has wave-like temperature oscillations ($Q\geq1$). The Q-factor is bound and its maximum value is obtained for $\tilde{\omega}=\pm\tilde{\omega}_{Qmax}=\pm Z^{-1/2}$, yielding:
\begin{equation}
Q(\pm\tilde{\omega}_{Qmax})=Z^{-1/2}.
\label{Q_omega_max}
\end{equation}

\noindent The curve $\tilde{\omega}_{Qmax}$ is plotted as a function of $Z$ in Figure  \ref{Q_zero_or_one} b as a red dashed-line.
The CV limit is restored for $Z\rightarrow 0$ where both $\tilde{\omega}_{Qmax}$ and $Q$ diverge.\\
For $Z\geq 1$ the wave-like temperature oscillations, although present, are overdamped ($Q\leq 1)$. 

\indent We now focus on the case in which underdamped wave-like temperature oscillations may be admitted, i.e. $0<Z<1$. The dispersion relation - $|\tilde{k}_{1}|$ (blue curve, left axis) and $|\tilde{k}_{2}|$ (red curve, right axis) - and the Q-factor are plotted vs $|\tilde{\omega}|$ in Figure \ref{complex_k_dispersion} c and d respectively, where a value of $Z=0.01$ has been chosen for sake of comparison with Section \ref{subsTauTminTauq_packetZ}.
Although the example illustrated in Figure \ref{complex_k_dispersion} d shows underdamped oscillatory solutions for all angular frequencies, a clear resonance in $Q$ stems out. This fact bears great relevance from an experimental stand point, where one seeks the greatest possible Q-factor. Practically the material thus behaves as an actual \textit{passband filter} for temperature oscillations both with respect to $\tilde{\omega}$,  as displayed in Figure \ref{complex_k_dispersion} d, and $\tilde{k}_{1}$, the latter assertion arises mapping $\tilde{k}_{1}$ vs $\tilde{\omega}$ (blue curve in Figure \ref{complex_k_dispersion} c) and plotting the $Q$-factor accordingly.\\

\begin{figure}[h!]
\begin{center}
\includegraphics[width=0.5\textwidth]{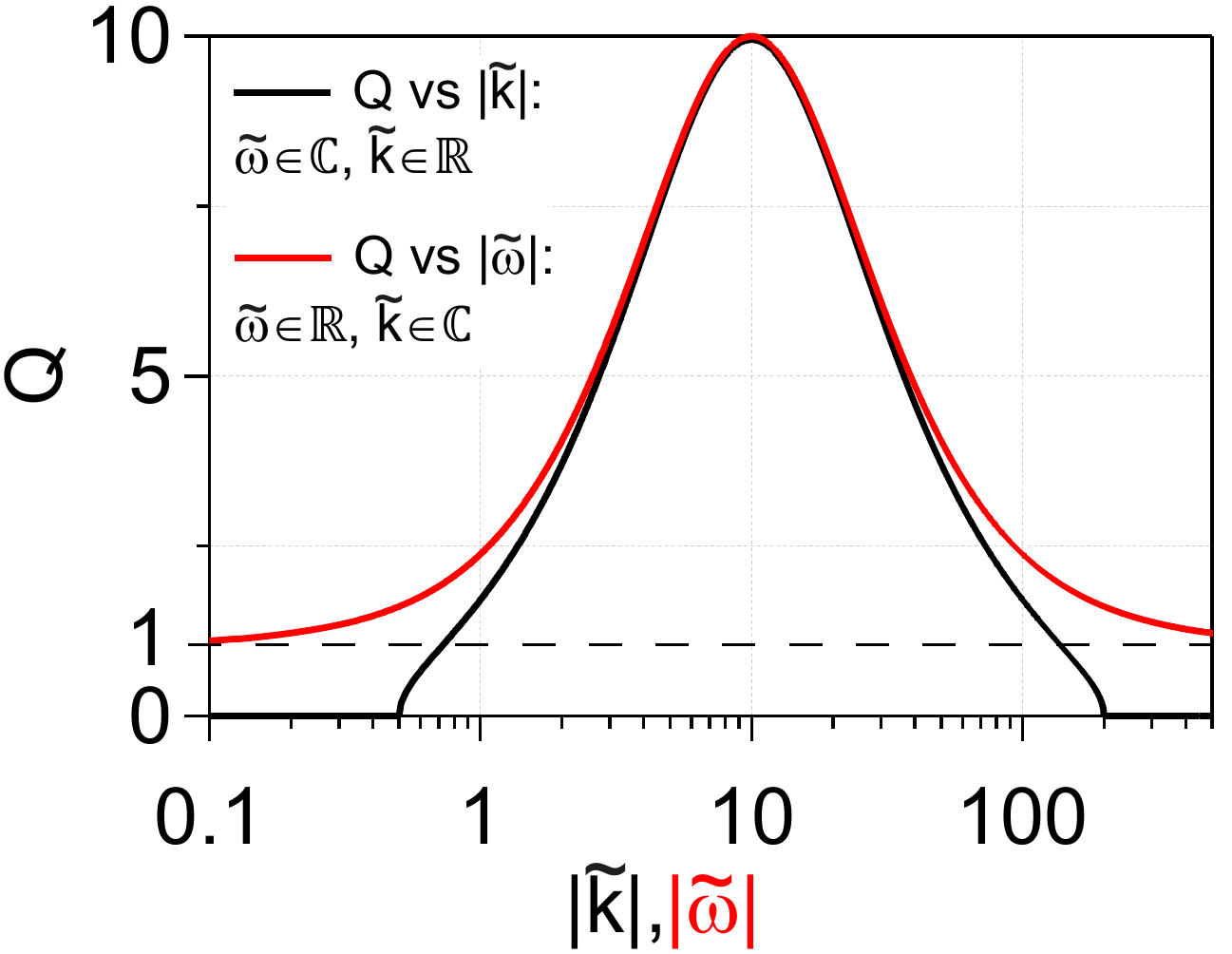} 
\caption{$Q$-factor vs $|\tilde{k}|$ for the case $\tilde{\omega}\in\mathbb{C}$ and $\tilde{k}\in\mathbb{R}$ (black curve).
$Q$ factor vs $|\tilde{\omega}|$ for the case $\tilde{k}\in\mathbb{C}$ and $\tilde{\omega}\in\mathbb{R}$ (red curve). The plots are obtained setting $Z=0.01$. $Q=1$ reference (black-dashed line) highlighting the transition between the underdamped and overdamped regimes.}
\label{Q_omega_k_sovrapposti}
\end{center}
\end{figure}
\section{Comparison between the two scenarios\label{Comparison_between_the_two_scenarios}}
Although the spatial temperature pulse propagation and the forced temperature oscillation in time are two essentially different problems, they are alike when damping is exiguous.
To substantiate this point, in Figure \ref{Q_omega_k_sovrapposti} we report, within the same graph and for a value of $Z$=0.01, the curves $Q$=$Q(\tilde{k})$, for the $\tilde{\omega}\in\mathbb{C}$, and $\tilde{k}\in\mathbb{R}$ case (black line) and $Q$=$Q(\tilde{\omega})$, for the $\tilde{k}\in\mathbb{C}$ and $\tilde{\omega}\in\mathbb{R}$ case (red line).
The oscillation behaviour is substantially the same for $Q\ge5$.
This may be rationalised noting that, for the case of negligible damping terms in Equation \ref{temperature_wave_equation_adimension}, Jeffrey's equation becomes a free wave equation that is symmetric in the adimensional space, $\xi$, and time, $\beta$, variables. In reciprocal space, these variables map into $\tilde{k}$ and $\tilde{\omega}$ respectively, which, for negligible damping, are real-valued.
With these prescriptions Equation \ref{complex_dispersion} yields $\tilde{\omega}_1 \sim \tilde{k}$ and a negligible $\tilde{\omega}_2$, for the spatial temperature pulse propagation, $\tilde{k}_1 \sim \tilde{\omega}$ and a negligible $\tilde{k}_2$, for the forced temperature oscillation in time. 

\begin{figure}[h!]
\begin{center}
\includegraphics[width=0.7\textwidth]{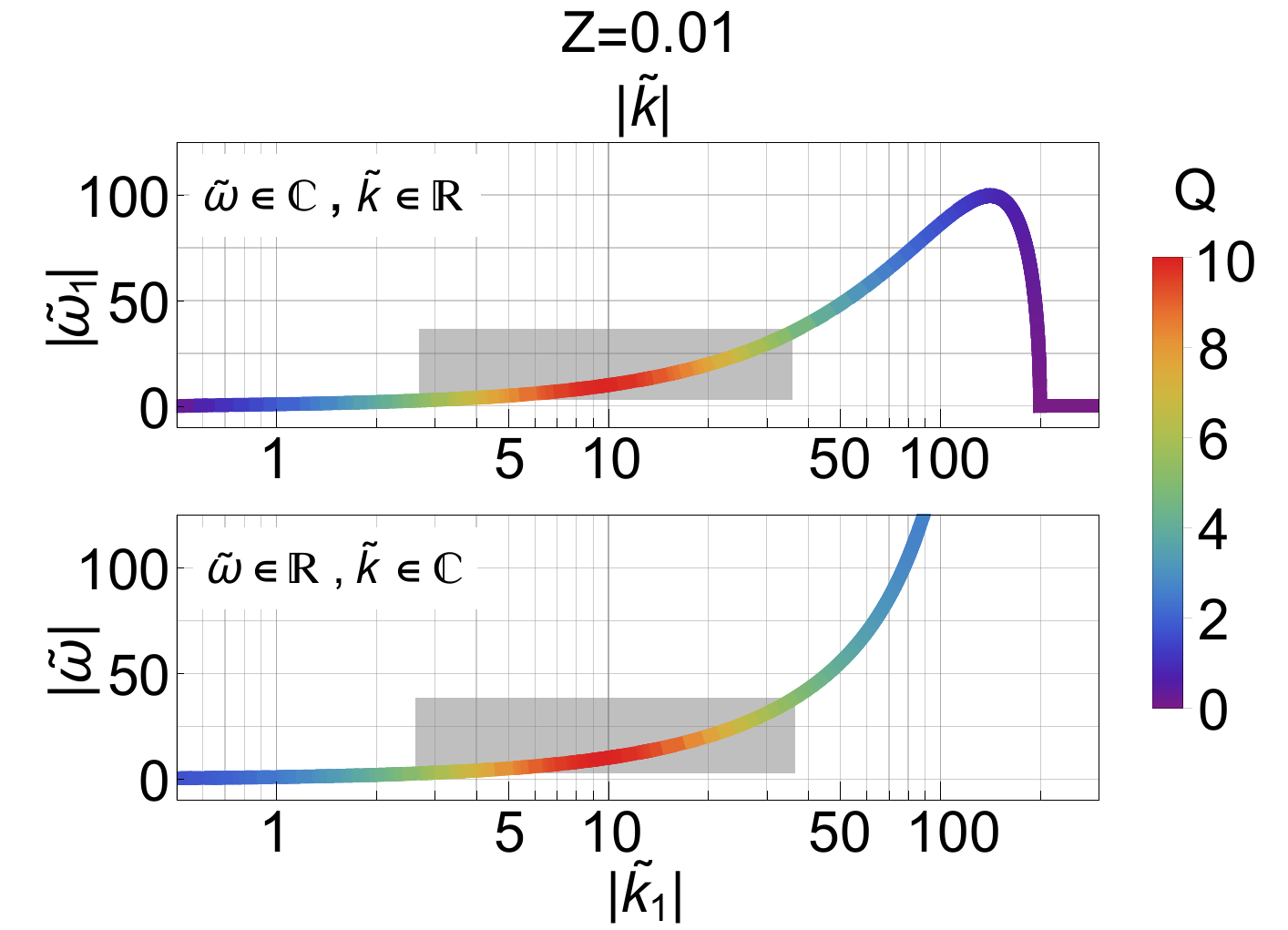} 
\caption{Dispersion relation for $Z=0.01$ in lin-log scale. Top panel: $|\tilde{\omega}_1|$ (vertical axis) vs $\tilde{|k|}$ (top horizontal axis) for the case $\tilde{\omega}\in\mathbb{C}$ and $\tilde{k}\in\mathbb{R}$. Bottom panels: $|\tilde{\omega}|$ (vertical axis) vs $|\tilde{k}_1|$ (bottom horizontal axis) for the case $\tilde{\omega}\in\mathbb{R}$ and $\tilde{k}\in\mathbb{C}$. The $Q$-factor is plotted in color scale. The shaded grey areas highlight the regions in which $Q\ge5$.}
\label{comparison}
\end{center}
\end{figure}

The two scenarios share, for high enough $Q$-factors, very similar dispersion relations. Figure \ref{comparison} reports the dispersion relations both for the spatial temperature pulse propagation (top panel), and the forced temperature oscillation in time (bottom panel), for a value of Z=0.01. The Q-factor is superposed on the dispersion relation as a color map. The two dispersion relations are substantially equal for the case of  $Q\ge5$, a region highlighted by the shaded grey area.

\section{Applications to real materials: case studies\label{Case_studies}}
In the following, we address the possibility of observing temperature wave-like oscillations in real systems based on the specific material's bandpass filter characteristics. Several systems are analysed following a top-down progression for the temperature oscillation dynamics both in time - from seconds to picoseconds - and space-scale - from millimetres to nanometers. The underlying microscopic physics differs substantially, the focus laying on the lattice, electronic or spin temperature. Previously reported observations of temperature oscillations in macroscopic granular media and solid He4 are first revised at the light of the present formulation. Next, we put forward predictions for bandpass filters characteristics of quantum materials as potential candidates for all-solid state thermal nanodevices, namely strongly correlated copper oxides and magnetically frustrated iridates. As a last case study, we revisit, at the light of the present formulation, the recent report of temperature oscillations in graphite at 80 K measured via transient thermal grating (TTG) spectroscopy \cite{huberman2019}. Graphite and other layered or 2D materials indeed constitute an alternative class of materials with potential for all-solid state thermal devices operating in the temperature wave-like regime \cite{cepellotti2015phonon}.
We here restore dimensional variables for comparison with real materials via the transformations $k=\tilde{k}/\sqrt{\alpha\tau_q}$ and  $\omega=\tilde{\omega}/\tau_q$.

We first focus on the spatial temperature pulse case. In order to observe temperature wave-like oscillations the thermal wave vector has to be comprised in the material's pass-band, $k\in[k_{Q=1,lo},k_{Q=1,hi}]$, while the corresponding thermal wavelength should exceed the system characteristic length $L$, $\lambda=2\pi /k\ge L$.
The ideal situation is the one where $\lambda_{Q_{max}}>L$, $\lambda_{Q_{max}}=2 \pi /k_{Q_{max}}$ being the thermal wavelength of the best oscillating mode. Practically, the relevant time-scale to observe oscillations is comprised between the oscillation period 2$\pi$/$\omega_{1}(k)$ and $Q(\omega_{1})$ times its value, the quality factor being a measure of the number of cycles it takes for an oscillation to die-off.

\begin{table}[t]
\centering
\begin{tabular}{|c|c|c|c|c|c|c|} 
 \hline  
&	Sand	&	Bio & Solid	&	COSCs   & MFI&Graphite\\
&  &tissue &He4 &&&\\  \hline
$\alpha \left[\um{m^2\over s}\right]$	&3$\times$10$^{-7}$& 1.5$\times$10$^{-7}$ &2&	6$\times$10$^{-7}$ &1$\times$10$^{-10}$&0.02 \\  \hline
$\tau_q\  [\um{s}]$	&8.9&$16$& 6$\times$10$^{-5}$&1$\times$10$^{-12}$ &1$\times$10$^{-9}$&1.8$\times$10$^{-9}$	\\ \hline
$\tau_T\  [\um{s}]$	&4.5&0.05&5$\times$10$^{-9}$&5$\times$10$^{-14}$&1$\times$10$^{-13}$&3$\times$10$^{-12}$\\   \hline
$Z$	&0.5&3$\times$10$^{-3}$&8$\times$10$^{-5}$&0.05&1$\times$10$^{-4}$&2$\times$10$^{-3}$\\  \hline
\end{tabular}
\caption{Thermal parameters for sand \cite{tzou2014macro}, biological living (Bio) tissue \cite{antaki2005new}, \textit{phonons} in solid He4 at 0.6 K and 54.2 atm \cite{tang1999wavy}, \textit{electrons} in Bi$_2$Sr$_2$CaCu$_2$O$_8$ (BiSCCO) at 20 K, \textit{spins} in the iridates Sr$_2$IrO$_4$ and Na$_2$IrO$_3$ \cite{gandolfi2017emergent} and \textit{phonons} in graphite at 80 K (refer to the main text for the extimation of the graphite thermal parameters). The values of BiSCCO and Sr$_2$IrO$_4$/Na$_2$IrO$_3$ are representative for the general cases of copper oxides superconductors (COSCs) and magnetically frustrated iridates (MFI) respectively. $Z$ values have been rounded to the first significant figure for sake of simplicity.}
\label{tableSummaryCoeffTemperature}
\end{table}

Granular materials have been proposed as possible systems sustaining temperature wave-like oscillations.
Although still a debated issue \cite{roetzel2006}, signatures of temperature wave-like oscillations were reported for instance in cast sand \cite{tzou2014macro} and biological living tissues considered as non-homogeneous fluid-saturated porous media \cite{Kuo_bioheat_heat_transfer,antaki2005new,zhang2009generalized,afrin2012numerical}.
Starting from the thermal parameters reported in Table \ref{tableSummaryCoeffTemperature}, we calculated the pass-band filter characteristics for the temperature pulse propagation, see Table \ref{omega_complex_table}.
For the case of sand, a value of $Z\sim0.5$ yields a $Q(k_{Qmax})\sim1$ for a thermal wavelength $\lambda_{Q_{max}}\sim$ 7 mm.
Although $\lambda_{Q_{max}}>L$, the characteristic dimension $L$ being the sand grain size ($L< 1$ mm), the filter pass-band width approaches zero making it hard to detect any signature of temperature oscillations. 
The situation is better off indeed for the case of biological living tissues.
$Z\sim10^{-3}$ allows for a fully developed pass-band of band-width (BW) $\sim$10$^{5}$ m$^{-1}$, where the BW is defined as BW=$k_{Q=1,hi}-k_{Q=1,lo}$.
A value of $Q(k_{Q_{max}})\sim20$ is obtained for $\lambda_{Q_{max}}\sim500$ $\um{\mu m}$, and $\lambda_{Q_{max}}\gg L$, where $L\sim$ few $\um{\mu m}$ is the living tissue pore dimension.
The optimal thermal oscillation angular frequency is $\omega_{1}(k_{Q_{max}})\sim1$ rad/sec, thus setting the relevant time-scale to observe oscillations in the range $\sim$ 1$-$100 seconds (i.e. in between the oscillation period and Q times its value).\\
\indent We now analyse the band-pass characteristics for the \textit{phonon} temperature of solid He4 at 0.6 K and 54.2 atm \cite{ackerman1968}.
The thermal parameters for the present case, see Table \ref{tableSummaryCoeffTemperature}, yield $Z\sim 8\times 10^{-5}$.
The bandpass filter has outstanding characteristics, see Table \ref{omega_complex_table}.
A $Q(k_{Q_{max}})\sim$ 100 is obtained for a $\lambda_{Q_{max}}\sim600$ $\um{\mu m}$, a value orders of magnitude in excess with respect to $L$= 0.3 nm, the unit cell dimension, the phonon thermal wavelength should not in fact be smaller than the minimum phonon wavelength.
The oscillation angular frequency $\omega_{1}(k_{Q_{max}})\sim$10$^{6}$ rad/sec yields a period of $\sim$ 3 $\um{\mu}$s, setting the time-scale for the observation of temperature oscillations in the range 3$-$300 $\um{\mu}$s (i.e. in between the oscillation period and Q times its value).
The BW $\sim10^6$ m$^{-1}$, exceeding $k_{Q_{max}}$ by almost two orders of magnitude, allows to excite thermal wavelengths down to the 10 $\um{\mu m}$ range with a Q-factor $\sim$ 3. However interesting under a scientific stand-point, He4 is not suited for potential thermal device applications.\\
\indent Turning to quantum materials, strongly-correlated oxides have recently been proposed as potential candidates to observe temperature wave-like oscillations via ultrafast optical techniques \cite{gandolfi2017emergent}.
The intrinsic anisotropy, together with strong correlations, grant a value of $Z\ll1/2$ at above liquid helium temperatures, see Table \ref{tableSummaryCoeffTemperature}.
Furthermore, these solid-state systems are amenable to nano-structuring.
These peculiarities makes them potential candidates for thermal device concepts based on temperature wave oscillations, operating on ultra-fast time scale and nanometer space-scale.
For instance Bi$_2$Sr$_2$CaCu$_2$O$_8$ (BiSCCO), the paradigmatic high-temperature superconductor, with a superconducting transition temperature as high as 100 K, behaves, at a lattice temperature of 20 K, as a passband for the \textit{electronic} temperature. The filter, which salient figures are summarised in Table \ref{omega_complex_table}, is characterised by a maximum $Q$-factor $\sim$ 4, indicating that the temperature oscillations are potentially accessible on the ultrashort time, 2$\pi/\omega_{1}(k_{Q_{max}}) \sim$ 1 ps, and space, $\lambda_{Q_{max}}\sim$ 1 nm, scales.
The value of $\lambda_{Q_{max}}$ is of the order of $L$, here taken as the crystal unit-cell dimension along the the $c$-axis.\\ 
\indent Magnetic materials are another class of quantum materials where these concepts apply and are foreseen to be fruitful in terms of applications. In this case the wave-like temperature propagation refers to the \textit{spin} temperature, which can decouple from the lattice and electronic temperatures in out-of-equilibrium conditions. Exploitation of coherent propagation of spin temperature opens interesting opportunities for spintronic-based nanodevices and for the use of magnetic materials exhibiting spontaneous magnetic nanotexturing.
We here address the case of magnetically frustrated iridates. These materials undergo an antiferromagnetic phase transition at a N\'eel temperature $T_{N}$, whereas, above $T_{N}$, maintain short range magnetic correlations up to a temperature $T_{corr}\gg T_{N}$.
For instance, Na$_2$IrO$_3$ undergoes a zig-zag magnetic transition at $T_{N}$=15 K , while short range correlations are retained at temparetures as high as $T_{corr}\sim$100 K \cite{hwan2015,nembrini2016}.
The characteristic $L$ is now dictated by the spin-spin correlation length. The interplay of the thermal parameters, see Table \ref{tableSummaryCoeffTemperature}, gives $Z$=10$^{-4}$, thus leading to potential oscillations.
Their passband filter characteristic, see Table \ref{omega_complex_table} for a paradigmatic example, is characterised by a $Q_{max}$$\sim$100 occurring at a $\lambda_{max}$$\sim$10 pm, a value much smaller than any physically sound $L$. Nevertheless, the BW is wide enough to allow achieving $Q$$\sim$2.5 for $\lambda$=1.5 nm, that is of the order of $L$ for the case of Na$_2$IrO$_3$.
The fact that $L$ decreases in the temperature range $T_N<T<T_{corr}$ allows to reduce $\lambda$ (increase $k$) so as to achieve higher $Q$-values at temperatures above $T_{N}$.
The exact scaling of the correlation length with temperature in these iridates is yet under investigation, nevertheless, assuming values for the coherence length down to 3 $\um{\AA}$ \cite{hwan2015}, results in a $\lambda$$\sim$$L$ yielding a $Q$ factor in excess of 10 with an oscillation angular frequency $\omega_{1}$$\sim$6$\times$10$^{9}$ rad/sec. The time scale to observe temperature oscillation thus falls in the 1$-$10 ns range (i.e. in between the oscillation period and $Q$ times its value) with thermal wavelengths in the nm range.
\begin{table} [h!]
\centering
\begin{tabular}{|c|c|c|c|c|c|c|}
 \hline
&	Sand	&Bio & Solid	&	COSCs   & MFI  &Graphite \\
&  &tissue &He4 &&&\\  \hline
$Q(k_{Qmax})$	&$	1	$&$	20	$&$	100	$&$	4	$&$100$&20\\  \hline
$k_{Qmax}$	&\multirow{2}{*}{$900$}&\multirow{2}{*}{1$\times$10$^{4}$}&\multirow{2}{*}{1$\times$10$^{4}$}&\multirow{2}{*}{6$\times$10$^{9}$}&\multirow{2}{*}{3$\times$10$^{11}$}&\multirow{2}{*}{4$\times$10$^{6}$}\\ 
$[1/\um{m}]$&&&&&&\\ \hline
$\lambda_{Qmax}[\um{m}]$	&$	0.007	$&	5$\times$10$^{-4}$&6$\times$10$^{-4}$&1$\times$10$^{-9}	$&2$\times$10$^{-11}$&2$\times$10$^{-6}$\\  \hline
$\omega_1(k_{Qmax})$ 	&\multirow{2}{*}{$0.1$}&\multirow{2}{*}{1}&\multirow{2}{*}{2$\times$10$^{6}$}&\multirow{2}{*}{4$\times$10$^{12}$}&\multirow{2}{*}{1$\times$10$^{11}$}&\multirow{2}{*}{1$\times$10$^{10}$}\\  
$ \left[\um{rad/ s}\right]$&&&&&&\\ \hline
BW $[1/\um{m}]$	&$	0	$&3$\times$10$^{5}$&2$\times$10$^{6}$&3$\times$10$^{10}$&4$\times$10$^{13}$&1$\times$10$^{8}$\\  \hline
\end{tabular}
\caption{Temperature wave bandpass filter salient characteristics for the $\omega\in\mathbb{C}$ and $k\in\mathbb{R}$ case, i.e. spatial temperature pulse:
best oscillating modes Q-factor, $Q(k_{Qmax})$, and corresponding wave vector $k_{Qmax}$, wavelength $\lambda_{Qmax}=2\pi/k_{Qmax}$, angular frequency $\omega_1(k_{Qmax})$ and filter bandwidth BW for sand, biological living (Bio) tissues, phonon temperature in solid He4 at 0.6 K and 54.2 atm, electronic temperature in Bi$_2$Sr$_2$CaCu$_2$O$_8$ (BiSCCO) at 20 K, spin temperature in the Sr$_2$IrO$_4$ and Na$_2$IrO$_3$ iridates and phonon temeprature in graphite at 80 K. The values of BiSCCO and Sr$_2$IrO$_4$/Na$_2$IrO$_3$ are representative for the general cases of copper oxides superconductors (COSCs) and magnetically frustrated iridates (MFI) respectively.
The values have been rounded to the first significant figure for sake of simplicity.
}
\label{omega_complex_table}
\end{table}

\begin{figure}
\begin{center}
\includegraphics[width=1\textwidth]{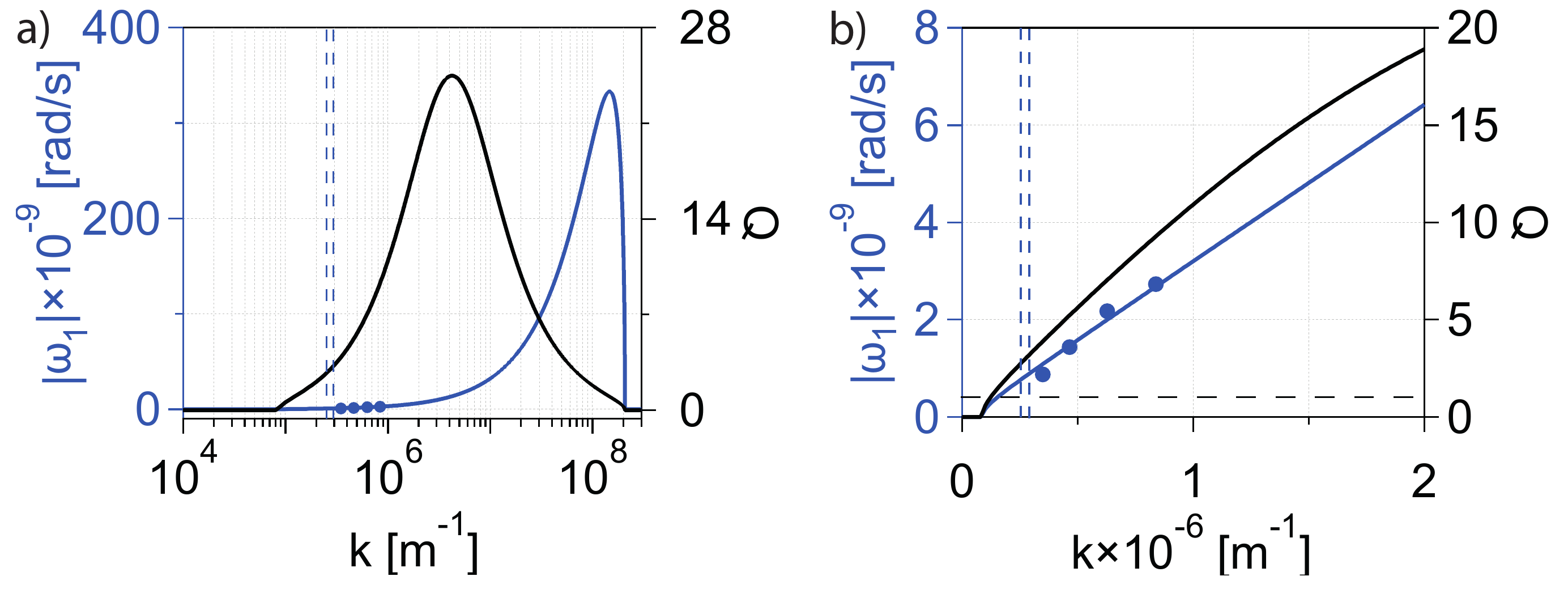} 
\caption{(a) $\omega_1$ (left axis, blue color) and Q-factor (right axis, black color) vs $k$ (horizontal axis) for the in-plane temperature oscillations in graphite at 80 K. The vertical axis are in lin scale, whereas the horizontal axis is in log scale. The full circles represent the oscillation angular frequencies measured by Huberman et al. \cite{huberman2019} with the TTG technique. The full line plots are calculated via Equation \ref{disp_tauTmintauq_DPL} (blu color) and Equation \ref{Q_factor_TauqmagTauT_DPL} (black color) (i.e. for the case $\tilde{\omega}\in\mathbb{C}$ and $\tilde{k}\in\mathbb{R}$) upon insertion of the optimal fitting parameters, $\tau_{Q}$=1.8 ns and $\tau_{T}$= 3 ps, to Huberman et al. data and resorting back to dimensional variables. The two vertical dashed blu lines indicate two additional $k$ values, in principle excited in Huberman et al. experiments, for which no temperature oscillations were actually detected. The $k$ range has been chosen to display the entirety of the theoretically predicted oscillatory modes. (b) Expanded view around the $k$ range probed in Huberman et al. experiment. For sake of visualization, the ratio of the maximum left and right axis range has been changed with respect to panel (a). The black horizontal dashed line highlights Q=1.}
\label{Graphite_graph}
\end{center}
\end{figure}

We now turn to the rationalisation, in the frame of the outlined theoretical approach, of $\textit{phonons}$ temperature oscillations in graphite at 80 K, as recently measured by Huberman et al. in their seminal work \cite{huberman2019} via the TTG technique. In a nutshell, two short laser pulses (temporal duration $\le$ 60 ps) were crossed at the surface of the sample, providing a transient spatially sinusoidal heat source of period $P$ set by the optical interference pattern. The explored periodicities were $P$=$\{$24.5, 21, 18, 13.5, 10, 7.5$\}$ $\um{\mu m}$. In so doing the authors launched temperature oscillations of thermal wave-vectors $k$=$2\pi/P$ and detected the corresponding oscillation angular frequencies $\omega_{1}$, as encoded in the time-dependent diffraction of a continuous-wave probe laser beam. The experiment thus falls in the $\tilde{\omega}\in\mathbb{C}$ and $\tilde{k}\in\mathbb{R}$ case. The experimental $\omega_{1}$ vs $k$ dispersion is reported as blue full circles in Figure \ref{Graphite_graph}. We pinpoint that no oscillations were reported for the two smaller $k$ values of 2.5$\times$10$^{-5}$ m$^{-1}$ and 3$\times$10$^{-5}$ m$^{-1}$, indicated by blue dashed vertical lines in Figure \ref{Graphite_graph}.

\begin{figure}
\begin{center}
\includegraphics[width=0.5\textwidth]{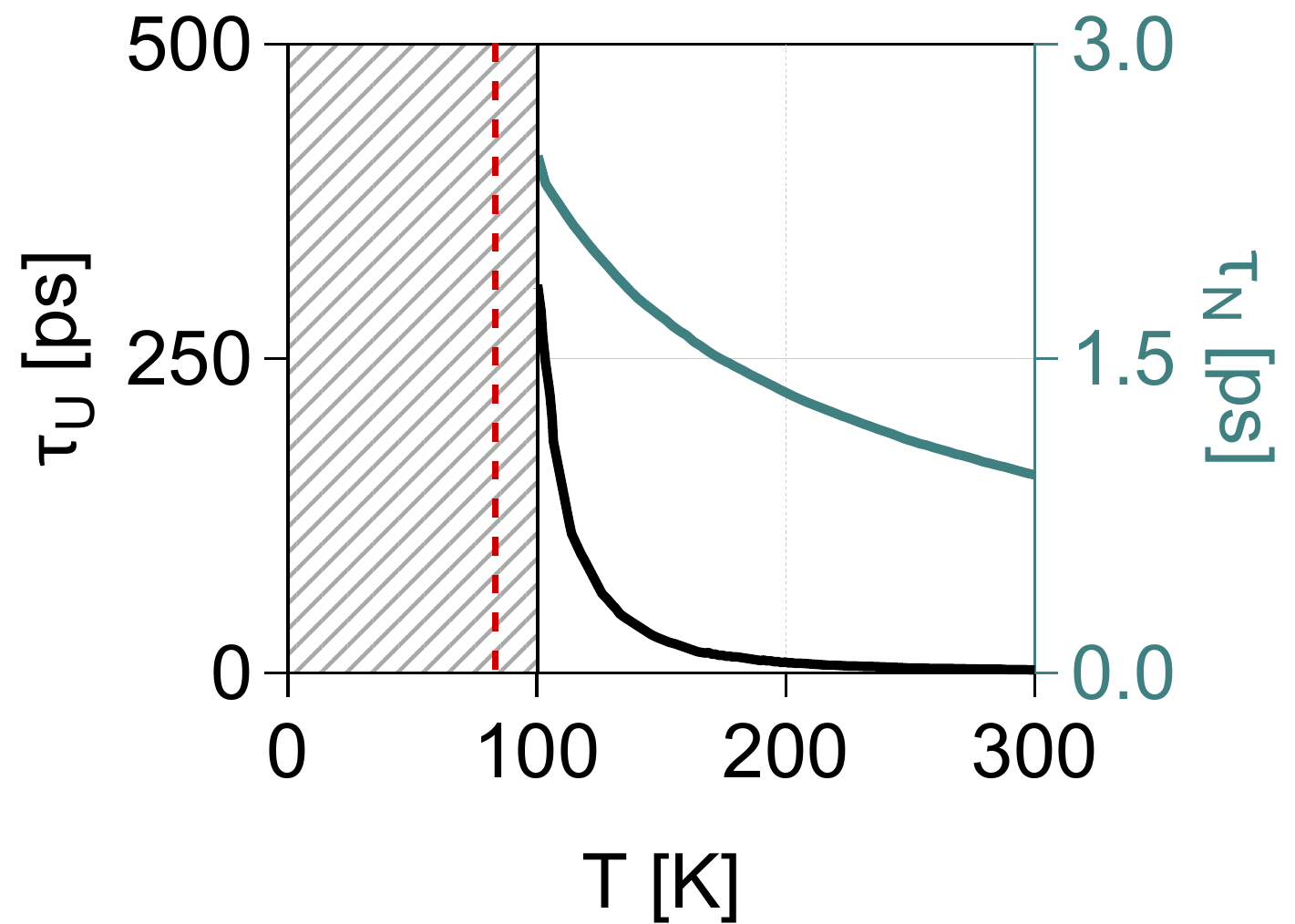} 
\caption{Average phonon scattering time for Umklapp (black curve, left axis) and Normal (emerald curve, right axis) processes as a function of the temperature (horizontal axis) in graphene. The plots are derived from the data reported in Figure 2 of Reference \cite{cepellotti2015phonon}, as discussed in the main text. The scattering times are not  available for temperature below 100 K (shaded region). The red dashed line indicates the temperature of 80 K, at which Huberman et al.\cite{huberman2019} reported observation of temperature waves.}
\label{U_N_scattering_time}
\end{center}
\end{figure}

We argue that, for graphite at cryogenic temperatures, the identifications $\tau_{T}$=$\tau_{N}$ and $\tau_{Q}$=$\tau_{U}$ hold, where $\tau_{N}$ and $\tau_{U}$ are the average phonon scattering times for Normal ($N$) and Umklapp ($U$) processes, respectively. Microscopically, $N$ processes lead to a momentum-conserving phonon distribution. This allows defining a local phonon temperature, hence, the onset of a temperature gradient at a time $\tau_{N}$ following the impulsive creation of the transient grating. $N$ processes do not contribute to heat transport, $U$ processes rather being responsible for it \cite{ziman1979,cepellotti2015phonon}. The onset of heat transport hence occurs at a time $\tau_{U}$ after the impulsive creation of the transient grating. As for an estimate of $\tau_{T}$ and $\tau_{Q}$, we first note that, in their experiment, Huberman et al. trigger in-plane heat transport in graphite. This fact legitimates, in the present context, exploitation of temperature dependent data available for graphene \cite{cepellotti2015phonon} as van der Waals  interactions, effective among graphite layers, do not drastically affect the lattice dynamics of individual graphene layers. Figure \ref{U_N_scattering_time} shows $\tau_{U}$ (black curve, left axis) and $\tau_{N}$ (emerald curve, right axis) in the temperature range 100-300 K, here estimated as $\tau_{N,U}$=1/$<\Gamma_{N,U}>$, where $<\Gamma_{N,U}>$ are the average line-widths, in Hz, for $U$ and $N$ processes reported in Figure 2, panel a) of Reference \cite{cepellotti2015phonon}. What emerges is that, around 100 K, $Z\ll$1/2, hence allowing for the observation of temperature wave-like behaviour. Despite the fact that the data for $\tau_{N}$ and $\tau_{U}$ are available down to 100 K only,  whereas Huberman et al. experiment is performed at 80 K, Figure \ref{U_N_scattering_time} clearly suggests that $Z\ll$1/2 should also hold at 80 K (vertical red-dashed line). In fact, as the temperature is lowered, $U$ scattering events become more rare, hence increasing $\tau_{U}$. Upon lowering the temperature in the whereabout of 100 K, the increase of $\tau_{U}$ is steeper with respect to the increase of $\tau_{N}$, allowing to foresee a further reduction of $Z$ at 80 K, thus resulting in an even better condition to observe temperature-wave like oscillations. 

Having shown that it is sound to rationalize temperature oscillations in graphite in the frame of our approach, we then fit the the experimental data of Huberman et al. via the $\omega_{1}$ vs $k$ dispersion given by Equation \ref{disp_tauTmintauq_DPL} (i.e. with the dispersion relation for the case $\tilde{\omega}\in\mathbb{C}$ and $\tilde{k}\in\mathbb{R}$) with dimensional variables restored and $\tau_{T}$ and $\tau_{Q}$ as fitting parameters. 
In the process of resuming dimensional variables, the thermal diffusivity at 80 K is set to\footnote{The value of $\alpha$ is obtained as the ratio between the in plane thermal conductivity for graphite, $k_{T}$=4300 W/mK \cite{fugallo2014thermal}, and its volumetric heat capacity $C$=2.35$\times$10$^{5}$ J/m$^3$K, both taken at 80 K. $C$ is calculated as $\rho$$C_{p}$, where $\rho$=2260 kg/m$^{3}$ \cite{pierson2012handbook} and $C_{p}$=104 J/kgK \cite{desorbo1953specific} are graphite's density and heat capacity per unit mass at 80 K, respectively.} $\alpha$=1.83$\times$10$^{-2}$ m$^{2}$/s. The best fit values are found to be $\tau_{T}$= 3 ps\ and $\tau_{Q}$=1.8 ns, which are lined up with the order of magnitude values that may be foreseen at 80 K by inspecting Figure \ref{U_N_scattering_time}, and yield $Z$=1.7$\times$10$^{-3}$.
The theoretical $\omega_{1}$ vs $k$ dispersion, with the optimal fit parameters inserted, is plotted as a full blue line (left axis, blue color) in Figure \ref{Graphite_graph}, panel a and b. In the same figure we also plot the $Q$ vs $k$ curve (right axis, black color), derived from Equation \ref{Q_factor_TauqmagTauT_DPL} upon insertion of the same delay times. The theoretical $\omega_{1}$ vs $k$ dispersion very well fits the experimental one. We emphasize that the optimal fit values $\tau_{T}$ and $\tau_{Q}$ are consistent with the values that may be expected assigning $\tau_{T}$=$\tau_{N}$ and $\tau_{Q}$=$\tau_{U}$ extrapolated at 80 K, see Figure \ref{U_N_scattering_time}. The calculated temperature wave group velocity is rather constant over the set of the experimentally explored $k$ values (see Equation \ref{group_velocity_Znon0} upon introduction of the optimal delay times) and reads $v_{g}$=3300 m/s, departing from the experimental value of $\sim$ 3200 m/s \cite{huberman2019} by only 3$\%$. Furthermore, the $Q$ vs $k$ dispersion also suggests why in Huberman et al. experiment no oscillations were reported for the two smaller $k$ values of 2.5$\times$10$^{5}$ m$^{-1}$ and 3$\times$10$^{5}$ m$^{-1}$, from now on addressed as ``dark'' modes, see blue dashed vertical lines in Figure \ref{Graphite_graph}. In the experimentally explored $k$ range, $Q$ is a monotonously increasing function of $k$, see Figure \ref{Graphite_graph}, panel b, making it easier to detect high $k$ values modes. For the two ``dark'' modes we find $Q(k=2.5\times$10$^{5}$ m$^{-1})=2.7$ and $Q(k=3\times$10$^{5}$ m$^{-1})=3.3$, as shown by the intersection of the vertical dashed blue lines with the $Q$-factor dispersion (black curve, right axis). Despite the fact that the former two modes still remain underdamped, the smaller $Q$ values might practically hinder, in conjunction with the TTG set-up sensitivity issues, their experimental detection.

Besides accounting  for Huberman et al. results, our analytical approach suggests the optimal conditions to observe temperature wave-like behaviour in TTG experiments. Specifically, the temperature pass-band filter characteristic for graphite at 80 K reaches its maximum value $Q_{max}$=25 for $k_{Q_{max}}$=4.3$\times$10$^{6}$ m$^{-1}$, see Figure \ref{Graphite_graph}, panel a (black curve right axis), corresponding to a thermal wavelength $\lambda_{Q_{max}}$=1.5 $\mu$m$\gg$$L$$\sim$0.3 nm, the system's characteristic length scale, $L$, being graphite's unit-cell dimension in the $ab$ plane. The angular frequency is $\omega_{1}$($k_{Q_{max}}$)=14$\times$10$^{9}$ rad/s, a value within reach of the detection capabilities of time-resolved spectroscopies. The time scale to observe the optimal temperature oscillation thus falls in the 0.4$-$10 ns range (i.e. in between the oscillation period 2$\pi$/$\omega_{1}$($k_{Q_{max}}$) and $Q_{max}$ times its value). In order to impulsively trigger the best oscillating temperature mode, a transient grating with $P$=2$\pi$/$k_{Q_{max}}$= 1.5 $\mu$m is thus required, a figure within reach of present TTG spectroscopy \cite{minnich2015PRB,bencivenga2019}. The above discussed thermal parameters and pass-band filter characteristics are summerized, rounded to the \textit{first significant figure}, in Table \ref{tableSummaryCoeffTemperature} and \ref{omega_complex_table}, respectively. Reasoning on the same footing, and upon inspection of Figure \ref{Graphite_graph} for $k$ values in excess of $k_{Q_{max}}$, but always within the band-pass filter BW, it emerges that it is possible to trigger sub-$\mu$m temperature wavelengths in the hypersonic frequency range.

This case study shows that our theoretical frame allows inspecting, by a simple and intuitive analytical mean, temperature-wave oscillations in graphite. The fact that our analysis was carried out inserting delay times as derived from a fitting procedure is actually accidental. The fit values are in fact compatible with the delay times that may be expected at 80 K, meaning that, if such delay times were actually available from first principle or experiments, our prediction would not have relied on any fitting parameter. All the same, temperature waves in graphene and other 2D technologically relevant materials could be tackled analysing their band-pass filters characteristics.

For the sake of theoretical comparison we report in Table \ref{k_complex_table} the filter characteristics also for the case of the forced temperature oscillation in time. A detailed analysis may be derived, \textit{mutatis-mutandis}, from the dispersion relation and $Q$-factor of the $k\in\mathbb{C}$ and $\omega\in\mathbb{R}$ case. The numbers for the cases characterised by an high enough Q-factor almost match the one reported in Table \ref{omega_complex_table}, confirming, in practical cases, the theoretical explanations objects of Figures \ref{Q_omega_k_sovrapposti} and \ref{comparison}, not so for oscillations with lower $Q$-factors where the response is indeed quite different as detailed throughout Section \ref{Temperature_Sustained_Wave-propagation} and \ref{Comparison_between_the_two_scenarios}.

\begin{table}[h!]
\centering
\begin{tabular}{|c|c|c|c|c|c|c|c|} 
 \hline  
&	Sand	&	Bio& Solid	&	COSCs   & MFI&Graphite \\
&  &tissue &He4 &&&\\ \hline
$Q(\omega_{Qmax})$	&1&20&	100&4&100&20\\ \hline
$k_1(\omega_{Qmax})$	 &\multirow{2}{*}{700}&\multirow{2}{*}{1$\times$10$^{4}$}&\multirow{2}{*}{1$\times$10$^{4}$}&\multirow{2}{*}{6$\times$10$^{9}$}&\multirow{2}{*}{3$\times$10$^{11}$}&\multirow{2}{*}{4$\times$10$^{6}$}\\
$[1/\um{m}]$&&&&&&\\ \hline
$\lambda_{Qmax}$	 $[\um{m}]$&9$\times$10$^{-3}$&6$\times$10$^{-4}$&6$\times$10$^{-4}$&1$\times$10$^{-9}	$&2$\times$10$^{-11}$&2$\times$10$^{-6}$	\\ \hline
$\omega_{Qmax}$	 &\multirow{2}{*}{$0.2$}&\multirow{2}{*}{$1$}&\multirow{2}{*}{2$\times$10$^{6}$}&\multirow{2}{*}{4$\times$10$^{12}$}&\multirow{2}{*}{1$\times$10$^{11}$}&\multirow{2}{*}{1$\times$10$^{10}$}\\ 
$[\um{rad/s}]$&&&&&&\\ \hline
\end{tabular}
\caption{Temperature wave bandpass filter salient characteristics for the $\omega\in\mathbb{R}$ and $k\in\mathbb{C}$ case, i.e. forced temperature oscillation in time:
best oscillating modes Q-factor, $Q(\omega_{Qmax})$, and corresponding wave vector $k_1(\omega_{Qmax})$, wavelength $\lambda_{Qmax}=2\pi/k_1(\omega_{Qmax})$ and angular frequency $\omega_{Qmax}$ for sand, biological living (Bio) tissues, phonon temperature in solid He4 at 0.6 K and 54.2 atm, electronic temperature in Bi$_2$Sr$_2$CaCu$_2$O$_8$ (BiSCCO) at 20 K, spin temperature in the Sr$_2$IrO$_4$ and Na$_2$IrO$_3$ iridates and phonon temeprature in graphite at 80 K. The values of BiSCCO and Sr$_2$IrO$_4$/Na$_2$IrO$_3$ are representative for the general cases of copper oxides superconductors (COSCs) and magnetically frustrated iridates (MFI) respectively.
The values have been rounded to the first significant figure for sake of simplicity.}
\label{k_complex_table}
\end{table}

We wind up this overview with an outlook on nanoscale granular materials. In perspective, nanogranual materials, schematized as two-phase composite media, may be good candidates where to apply the present formalism seeking for temperature oscillations \cite{AICHLMAYR2006377, PETERSON2006257}. Specifically, the value of $Z$ may be tailored tuning the density and thermal conductivity of each phase, the effective cross thermal conductivity of the two phases and their volume fractions \cite{WANG20081751}. Control of these parameters is within reach of current technology. For instance, gas phase deposition allows achieving nanoporous scaffolds \cite{peli2016, benetti2017} with tailored volume fraction \cite{benetti2018}. As for the densities and thermal parameters they may be tailored by engineering the materials to be deposited \cite{benetti2019} and/or tuned by infiltrating the porous nanoscaffold with fluids \cite{benetti2017}.

\section{\label{Conclusion}Conclusion}
This work provides a straightforward, easy-to-adopt, analytical means to inspect the optimal conditions to observe temperature wave oscillations. The theoretical frame relies on the macroscopic DPL model in its first-order formulation. It parallels the approach successfully employed in solid state physics and optics to investigate electronic wavefunction and electromagnetic wave propagation in solid state devices. 
The complex-valued dispersion relation is investigated for the cases of a localised temperature pulse in space and of a forced temperature oscillation in time, respectively. A modal quality factor is introduced as the key parameter to access the temperature propagation regime. For the case of the temperature gradient preceding the heat flux, the quality factor allows mimicking the material as a frequency and wavelength filter for the temperature wave. The bandpass filter characteristics are achieved in terms of the relevant delay times entering the DPL model. Previous reports of temperature oscillations, arising in different physical contexts, are here revised at the light of the present formulation. Furthermore, the possibility of observing temperature wave-like oscillations in quantum materials at the nanoscale and on the ultra-fast time scale is addressed, based on the specific material's bandpass filter characteristics for temperature oscillations. Engineering temperature wave-like propagation in solid-state condensates and spin-temperature oscillations in magnetic materials will open the way to all-solid state thermal nanodevices, operating well above liquid helium temperature and tacking advantage of a wealth of excitations - i.e electrons, phonons and spins - and, possibly, of their mutual interplay. As a timely case study, recent experimental evidence of wave-like temperature oscillations in graphite, as measured via thermal transient gratings spectroscopy, is rationalized based on the graphite bandpass filter traits for temperature oscillations, the thermal wavelength and time period spanning in the $\mu$m to sub-$\mu$m and ns range, respectively. This adds a tool for the investigation of temperature wave propagation in graphine and other technologically relevant 2D materials.

The same approach can be extended \textit{mutatis-mutandis} to mass transport in the frame of the generalised Fick's law stemming from the first-order DPL model. With the due substitutions, the present results remain in fact valid with respect to mass density wave-like oscillations. 

The present formulation allows investigating, on the same footing, systems with relevant time-scales spanning from seconds to hundreds of femtoseconds and space scales ranging from millimetres to nanometers. This work will hence be beneficial toward designing thermal devices architectures where temperature waves may play a role as, for instance, in heat-spreading materials technology and ultra-fast laser assisted processing of advanced materials.

\section*{Aknowledgements}
Francesco Banfi acknowledges financial support from Universit\'e de Lyon in the frame of the IDEXLYON Project -Programme Investissements d' Avenir (ANR-16-IDEX-0005). Francesco Banfi and Marco Gandolfi acknowledge financial support from the MIUR Futuro in ricerca Grant in the frame of the ULTRANANO Project (Project No. RBFR13NEA4). Christ Glorieux and Marco Gandolfi are grateful to KU Leuven Research Council for financial support (C14/16/063 OPTIPROBE). Claudio Giannetti acknowledges support from Universit{\`a} Cattolica del Sacro Cuore through D.2.2 and D.3.1 grants.

\bibliography{Gandolfi_references}

\end{document}